\documentclass[letterpaper]{article}
\usepackage[english]{babel}
\usepackage[utf8x]{inputenc}
\usepackage{natbib}
\usepackage[T1]{fontenc}
\usepackage{amsmath,mathtools}
\usepackage{dsfont}
\usepackage[table]{xcolor}
\usepackage{setspace,caption}
\usepackage{lipsum}
\usepackage{tabularray}
\usepackage{adjustbox}

\usepackage[margin=1in]{geometry}

\usepackage{natbib}

\newcommand{\N}{\mathbb{N}}

\usepackage{amsmath}
\usepackage{amsfonts}
\usepackage{graphicx}
\usepackage[colorinlistoftodos]{todonotes}
\usepackage[colorlinks=true, allcolors=blue]{hyperref}
\usepackage{subcaption}
\usepackage{booktabs}
\usepackage{longtable}
\usepackage{multirow}
\usepackage{threeparttable}
\usepackage{threeparttablex}
\usepackage[section]{placeins}
\usepackage{dsfont}
\usepackage{float}
\usepackage{tikz}
\usepackage{amsthm}
\usetikzlibrary{arrows,chains,matrix,positioning,scopes}
\makeatletter
\renewcommand*\env@matrix[1][*\c@MaxMatrixCols c]{%
  \hskip -\arraycolsep
  \let\@ifnextchar\new@ifnextchar
  \array{#1}}
\makeatother

\theoremstyle{definition}

\newtheorem*{proposition*}{Proposition}

\newtheorem*{definition*}{Definition}

\newtheorem*{hypothesis*}{Hypothesis}

\title{Measuring Higher-Order Rationality with Belief Control\thanks{We thank 
Marina Agranov, Colin F. Camerer, John Duffy,
David Gill, Paul J.\@ Healy, Kirby Nielsen,
Thomas R. Palfrey, Joseph Tao-yi Wang and audiences at California Institute of Technology, University of Cologne,
the 2021 ESA North American Conference and the 2023 Los Angeles Experiments (LAX) Workshop for comments. 
We are grateful to the editors Roberto Weber and 
Ido Erev and two referees for their detailed and 
constructive comments.
We thank Wei-Ming Fu for his excellent research assistance. We also thank Joseph Tao-yi Wang
for his generosity and support of our use of Taiwan Social Sciences Experimental Laboratory (TASSEL)
at National Taiwan University. This study is approved by National Taiwan University 
IRB (201910HM015). The experimental design and analysis plan are 
pre-registered on the Open Science Framework (https://osf.io/gye4u/).
The data, analysis code, experimental instructions, and program are also available on the 
Open Science Framework (https://osf.io/neqxm/).
Wei James Chen is supported by Ministry of Science and Technology of Taiwan (MOST 109-2636-H-008-002). 
Po-Hsuan Lin is supported by National Science Foundation
(SES-2243268).
All errors are our own.}}
\author{
  Wei James Chen\thanks{Department of Agricultural Economics, National Taiwan University, 
  Taipei, Taiwan, 10617. ORCID: 0000-0001-6360-9348. Email: jamesweichen@ntu.edu.tw}\\
  National Taiwan University
  \and
  Meng-Jhang Fong\thanks{(Corresponding Author) 
  Amazon Science, Seattle, WA 98109 USA. ORCID: 0009-0008-8832-3985. Email: mengjhangfong@gmail.com}\\
  Amazon Science
  \and
  Po-Hsuan Lin\thanks{Department of Economics, University
  of Virginia, Charlottesville, VA, USA, 22904. ORCID: 0000-0003-3437-1734. Email: plin@virginia.edu}\\
    University of Virginia
}
\date{\today}

\begin{document}

\maketitle

\begin{abstract}
Determining an individual's strategic reasoning capability based solely on choice data is a complex task. 
This complexity arises because sophisticated players might have non-equilibrium beliefs about others, leading to non-equilibrium actions. 
In our study, we pair human participants with computer players known to be fully rational. 
This use of robot players allows us to disentangle limited reasoning capacity from belief formation and social biases. 
Our results show that, when paired with robots, subjects consistently demonstrate higher levels of rationality and maintain stable rationality levels across different games compared to when paired with humans. 
This suggests that strategic reasoning might indeed be a consistent trait in individuals. 
Furthermore, the identified rationality limits could serve as a measure for evaluating an individual's strategic capacity when their beliefs about others are adequately controlled.
\end{abstract}

\bigskip



\newpage
\section{Introduction}

Understanding whether individuals make optimal choices in strategic environments is a fundamental question in economics. 
Unlike individual decision-making, a game involves multiple players whose payoffs depend on each other's choice. 
In this setting, achieving equilibrium requires a player to exhibit both first-order rationality and higher-order rationality. 
This necessitates that players are not merely rational themselves but also operate under the assumption that their counterparts are rational. 
Furthermore, they must believe that other participants consider them to be rational, and this belief cascades infinitely. 
As a result, in equilibrium, each player's assumptions about the strategies of their peers match the actual strategies employed, allowing them to optimally respond.

However,  expecting players to engage in iterative reasoning and demonstrate infinite levels of rationality is notably demanding, especially when viewed empirically. 
This is evidenced by well-documented instances of players diverging from 
equilibrium play (see, for example, \citealt{camerer2003behavioral}).
Given these empirical discrepancies, a significant volume of research has been dedicated to determining the extent of iterative reasoning an individual can realistically execute within different contexts.

Apart from exploring the extent of iterative reasoning an individual can undertake, this paper delves into another crucial, related query: Is there consistency in an individual's depth of strategic reasoning across various games?
Measuring strategic reasoning abilities of interacting individuals can facilitate our understanding and predictions of individuals' 
behavioral patterns. It also helps us evaluate whether the observed non-equilibrium actions are driven by bounded rationality or by other factors. 
Nevertheless, if we observe no regularity when measuring one's depth of strategic reasoning in different environments,
there may not even exist such a persistent trait called ``strategic thinking ability.''

The main challenge behind inferring individual strategic reasoning ability from choice data is that the strategic sophistication revealed by one's choices does not directly imply the maximum steps of iterative reasoning one is able to perform. 
As noted by \cite{ohtsubo2006depth},\footnote{``Subjects who go through several levels of reasoning and figure out the equilibrium solution to the game, will in general not invoke the maximum depth of reasoning precisely because they do not assume---and perhaps should not assume---that the other $n-1$
players are as smart as they are'' (\citealp{ohtsubo2006depth}, p.\@ 45).} a player's observed depth of reasoning is determined not only by their reasoning capability but also by their beliefs about the opponents' (revealed) sophistication, a notion supported by empirical evidence in \cite{agranov2012beliefs} and \cite{alaoui2015endogenous}. 
An individual who can carry out more than $k$ steps of reasoning would act as a $k$th-order rational player when they believe that their
opponent exhibits $(k-1)$th-order rationality. 
In other words, measuring an individual's revealed strategic sophistication only yields a lower-bound estimate of their actual sophistication. 
In addition, psychological factors other than bounded rationality such as lying aversion and fairness concern may also motivate a player to deviate from an equilibrium \citep{cooper2016other}.
Without controlling for a player's beliefs and social preferences, the estimation of
their strategic reasoning ability could be unstable and lack external validity.

In a study on bounded strategic sophistication by \cite{georganas2015persistence}, a question similar to the one posed in this paper is explored. 
In their research, 
participants play two distinct families of games.
Although their study does not extensively control for participants' beliefs, it reveals a limited persistence of individual strategic sophistication between the two families of games.\footnote{Another study that reports inconsistent depth of reasoning across games is \cite{cooper2023consistent}, which examines the comparative statics predictions of the level-$k$ model without controlling for participants' beliefs. Note that the idea of examining cross-game persistence of reasoning depth can be traced back to \cite{stahl1995players}, who found that 72\% of subjects had a stable depth of reasoning, though they focused on a \textit{single} family of games.}

In this paper, we demonstrate a method to test the stability of individual strategic sophistication and to possibly pin 
down the upper bound of an individual's depth of strategic reasoning in the lab: having human subjects interact 
with equilibrium-type \textit{computer} players induced by infinite order of rationality. 
By informing human players that they are facing fully rational computer players, we are able to unify players' expectations about their opponents. 
Additionally, introducing computer players precludes the possible effect of social preferences \citep{houser2002revisiting, johnson2002detecting, van2008value}. 
Thus, human players with an infinite order of rationality are expected to select an equilibrium strategy. 
In this setting, out-of-equilibrium actions would provide us a solid ground to identify an individual's order of rationality for 
inferring their strategic reasoning ability since those actions are likely driven by bounded rationality.

To investigate the stability of individual strategic sophistication across games, we conduct an experiment with two classes of dominance-solvable games, ring games and guessing games. 
Proposed by \cite{kneeland2015identifying} for identifying higher-order rationality, an $n$-player ring game can be characterized by $n$ payoff matrices and has the following ring structure: the $k$th player's payoff is determined by the $k$th player's and $(k+1)$th player's actions, and the payoff of the last ($n$th) player, who has a strictly dominant strategy, is determined by the last and the first player's actions. 
We employ guessing games that represent a symmetric variant of the two-person guessing games studied by \cite{costa2006cognition}, in which a player's payoff is single-peaked and maximized if the player's guess equals its opponent's guess times a predetermined number.\footnote{The guessing game we implement in this paper diverges from the standard beauty contest game, primarily because the standard beauty contest game is not strictly dominant solvable. However, it is worth noting that if the beauty contest game involves only two players, then it becomes dominant solvable \citep{grosskopf2008two, Chen2017}.}

Among the games that have been used to study strategic reasoning, we choose to implement ring games and guessing games in our experiment for two reasons.  
First, our instruction of a fully rational computer player's behavior is tailored to align with the payoff structure of dominance-solvable games, in which the computer players' actions can be unambiguously determined (see Section \ref{subsec:treatments} for details).
Furthermore, these dominance-solvable games enable a structure-free identification approach, leveraging the notion of rationalizable strategy sets \citep{bernheim1984rationalizable, pearce1984rationalizable}. 
The core idea behind this identification approach is that, within a dominance solvable game, we can gauge an individual's depth of reasoning by assessing how many rounds of iterated deletion of dominated strategies the individual's chosen action would survive. 
Importantly, this approach does not impose structural assumptions on (the beliefs about) non-rational players' behavior. 
Therefore, these classes of games provide a plausible, structure-free method to empirically categorize individuals into distinct levels of rationality. 

Second, we intend to implement two types of games that are sufficiently different so that, if we observe any stability in individual strategic reasoning levels across games, the stability does not result from the similarity between games. 
We believe that ring games and guessing games are dissimilar to each other. 
On the one hand, a ring game is a four-player discrete game presented in matrix forms. 
On the other hand, a guessing game is a two-player game with a large strategy space, which is more like a continuous game. 
In fact, \cite{cerigioni2019higher} report that the correlation of their experimental subjects' reasoning levels between ring games and beauty contest games
is only 0.10.
Although not intended to provide conclusive evidence from a limited number of games, we believe our study takes an important step toward investigating the consistency of reasoning levels across diverse game types, in line with recent literature encouraging further examination of cross-game stability.

Our experiment comprises two treatments within each game family: the Robot Treatment and the History Treatment. 
In the Robot Treatment, subjects encounter computer players employing equilibrium strategies. 
In the History Treatment, subjects confront choice data from human players in the Robot Treatment.  
The History Treatment simulates an environment where human subjects interact without displaying social preferences and serves two main objectives.
First, by examining if a subject's observed rationality level in the Robot Treatment exceeds that in the History Treatment, we can evaluate whether the subject responds to equilibrium-type computer players by employing a strategy that reaches {their} full capacity for strategic reasoning.
Second, by comparing the individual orders of rationality inferred from data in both the Robot and History Treatments, we can investigate whether the introduction of robot players contributes to stabilizing observed rationality levels across various games.

Overall, our findings indicate that strategic reasoning ability may be a persistent personality trait deducible from choice data when subjects interact with robot players in strategic scenarios.
Relative to interactions with human opponents, we observe a larger proportion of participants adopting equilibrium strategies and demonstrating higher levels of rationality. 
This observation is supported by both our between- and within-subject statistical analyses, underscoring the effectiveness of our Robot Treatment and implying that the rationality levels exhibited in this treatment potentially approach subjects' strategic thinking capacity.\footnote{One might doubt if a subject has the motivation to act rationally upon the presence of an opponent with a (much) higher rationality level than the subject has. In Section \ref{subsection:robot_validity}, we argue that a subject does have the incentive to exhibit the highest order of rationality they can achieve when they know their opponent is at least as rational as themselves.}

Furthermore, our investigation reveals that subjects' rationality levels remain remarkably stable across distinct game classes when interacting with robot players. 
In terms of absolute levels, a substantial number of first-order and fourth-order rational players retain their respective types while transitioning from ring games to guessing games.
In the Robot Treatment, approximately 38\% of subjects exhibit constant rationality levels across games.\footnote{The constant rationality level
hypothesis is the strictest requirement for the stability of rationality 
levels across games. In \ref{appendix:additional_results}, we explore 
two weaker notions of stability and find that players' rationality 
levels are more stable across games in the Robot Treatment,
even when considering these weaker notions of stability.}
A further statistical test involving 10,000 simulated samples demonstrates that this stability in rationality levels cannot be attributed to two independent type distributions, with the actual proportion of constant-level players exceeding the mean simulated proportion by 6 percentage points. 
Additionally, applying the same statistical analysis to the History Treatment reveals no significant disparities in the proportions of constant-level players between actual and simulated datasets. 
This indicates that the stability in individual rationality levels is not solely due to game selection but is influenced by our manipulation of subjects' beliefs about opponents' depths of reasoning.

A subject's performance in other cognitive tests could potentially hold predictive power regarding their strategic reasoning performance in games. 
As such, we incorporate tasks measuring cognitive reflection, short-term memory, and backward induction abilities (see Section \ref{subsec:cognitive_tests} for details) into our experiment.
We observe that a subject's cognitive reflection and backward induction abilities are positively correlated with {their} levels of 
rationality, whereas no significant correlation is found with {their} short-term memory capacity.

The rest of the paper proceeds as follows. 
The next subsection reviews the related literature. 
Section \ref{sec:theory} summarizes the theoretical framework upon which our identification approach and hypotheses to be tested are based. 
Section \ref{sec:games} describes the ring games and guessing games implemented in our experiment. 
Section \ref{sec:identification} discusses how we identify a subject's 
rationality {level} given choice data. 
Section \ref{sec:experiment_design} presents our experimental design and the hypotheses to be tested. The experimental results are reported in 
Section \ref{sec:results}. In Section \ref{sec:discussion_new}, we
discuss the validity and limitations 
of our belief control approach for the Robot Treatment. Finally,
Section \ref{sec:conclusion} concludes. 
The complete instructions of our experiment can be found in Supplementary Information.\footnote{The provided instructions are originally in Chinese and have been translated into English.}

\subsection{Related Literature}

Over the past thirty years, various researchers have theoretically studied the idea of limited depth of reasoning, including \cite{selten1991anticipatory, selten1998features}, \cite{aumann1992irrationality}, 
\cite{stahl1993evolution}, \cite{alaoui2015endogenous, alaoui2022cost}, \cite{lin2022multicognitive}, and \cite{lin2024cognitive}. 
In addition to theoretical contributions, \cite{nagel1995unraveling} 
conduct the first experiment on beauty contest games and introduce the level-$k$ model to 
describe non-equilibrium behavior. The level-$k$ behavior has subsequently been observed in a variety of game types, including matrix games (e.g., \citealp{stahl1994experimental,stahl1995players, costa2001cognition, crawford2007fatal}), investment games (e.g., \citealp{rapoport2000mixed}), guessing games (e.g., \citealp{costa2006cognition}), undercutting games (e.g., \citealp{arad201211}), auctions (e.g., \citealp{crawford2007level}), and sender-receiver games (e.g., \citealp{cai2006overcommunication, wang2010pinocchio, fong2023extreme}), though the list is not exhaustive.

Unlike the literature that primarily investigates individuals' strategic sophistication 
within the context of a single specific game, our work, which is closely related to \cite{georganas2015persistence} (hereinafter, GHW), 
centers on the examination of the consistency of strategic sophistication across different games.
In particular, we follow the language of GHW to formalize our hypotheses to be tested.\footnote{For a brief summary of the model in GHW, see Section \ref{sec:model_in_georganas}; also, see Section \ref{subsec:hypotheses} for the hypotheses.} 
Although both GHW and this paper experimentally investigate whether a 
subject's sophistication type persists across games, our study differs from GHW in several ways. First, 
we substitute the ring games for the undercutting games in GHW and use a simplified, symmetric version of the guessing games. 
Second, we employ an identification strategy distinct from the standard level-$k$ model to determine a subject's strategic sophistication. 
We use dominance solvable games in order to identify higher-order rationality without imposing strong and ad hoc assumptions on players' first-order beliefs, which can in turn reduce the noise in the estimation of individual reasoning depth using a level-$k$ model.\footnote{\cite{burchardi2014out} conduct an experiment in a standard beauty contest with belief elicitation, finding heterogeneity in both level-0 beliefs and level-0 actions within a game.} 
More importantly, we control for human subjects' beliefs about opponents' sophistication (and social preferences) using computer players. 
As a result, we observe a higher correlation in subjects' types across games compared to GHW, in which subjects are matched with each other.

Ring games, first utilized for identifying higher-order rationality by \cite{kneeland2015identifying}, are subsequently studied by \cite{lim2016identifying} and \cite{cerigioni2019higher}, who investigate two variants of the ring games.
In this study, we follow the \textit{revealed rationality approach} adopted by \cite{lim2016identifying} and \cite{cerigioni2019higher} as our identification approach (discussed in Section \ref{sec:identification}). 
It is worth noting that \cite{cerigioni2019higher} also find little correlation in subjects' estimated types across various games, including ring games, e-ring games, $p$-beauty contests, and a $4 \times 4$ matrix game. 
Again, our results suggest that the lack of persistence in the identified order of rationality at the individual level is driven by subjects' heterogeneous beliefs about the rationality of their opponents. 

Indeed, several empirical studies have shown that beliefs about others' cognitive capacity for strategic thinking can alter a player's strategy formation. 
\cite{friedenberg2018bounded} indicate that some non-equilibrium players observed in the ring games \citep{kneeland2015identifying} may actually possess high cognitive abilities but follow an irrational behavioral model to reason about others. 
Alternatively, \cite{agranov2012beliefs}, \cite{alaoui2015endogenous}, \cite{gill2016cognitive}, and \cite{fe2022cognitive} find that, in their experiments, subjects' strategic behavior is responsive to the information they receive about their opponents' strategic abilities.\footnote{In \cite{agranov2012beliefs}, subjects play against each other, graduate students from NYU Economics Department, or players taking uniformly random actions. In \cite{alaoui2015endogenous}, subjects play against opponents majoring in humanities, majoring in math and sciences, getting a relatively high score, or getting a low score in a comprehension test. In \cite{gill2016cognitive} and \cite{fe2022cognitive}, subjects play against opponents with similar or differing performance in cognitive tests.}
The designs of experiments allow them to manipulate subjects' beliefs, whereas we aim to elicit and identify individual strategic capability by unifying subjects' beliefs about opponents.

Some recent studies have tried to distinguish between non-equilibrium players who are limited by their reasoning abilities and players who are driven by beliefs. 
Identifying the existence of ability-bounded players is important since, if non-equilibrium behavior is purely driven by beliefs, it would be unnecessary to measure an individual's reasoning depth. 
\cite{jin2021does} utilizes a sequential version of ring games, finding that around half of the second-order and third-order rational players are bounded by ability. 
\cite{alaoui2020reasoning} also report the presence of ability-bounded subjects by showing that an elaboration on the equilibrium strategy shifts the subjects' level-\textit{k} types toward higher levels. 
Overall, the existence of both ability-bounded and belief-driven players in the real world indicates the need for an approach that can measure individual reasoning ability without the impact of beliefs.
Whereas \cite{jin2021does} and \cite{alaoui2020reasoning} do not pin down the belief-driven players' actual ability limit, we aim to directly measure each subject's strategic ability.

\cite{bosch2020one} propose an approach to test a subject's reasoning level in a given game: letting a subject play against herself (i.e., an ``one-person'' game). 
Specifically, in their study, each subject acts as both players in a 
modified two-person \textit{p}-beauty contest \citep{grosskopf2008two, Chen2017}, in which a player's payoff decreases in the distance between their own guess and the average guess multiplied by $p$, and the subject receives the sum of the two players' payoffs.\footnote{\cite{bosch2020one} report that 69\% of the subjects do not select the equilibrium action (0, 0) when playing the one-person game, which echoes the findings of the presence of ability-bounded players in \cite{jin2021does} and \cite{alaoui2020reasoning}.} 
The one-person game approach eliminates the impact of beliefs that arises from interacting with human players. 
However, a limitation of this approach is that it can only be applied to the game in which the equilibrium is Pareto optimal. 
For instance, it would be rational for a payoff-maximizing subject to deviate from the equilibrium and choose (Cooperate, Cooperate) in the prisoner's dilemma since (Cooperate, Cooperate) maximizes the total payoff of both players even though those are not equilibrium strategies.\footnote{Also note that in the ring game G1, both the equilibrium strategy profile (P1: \textit{b}, P2: \textit{c}, P3: \textit{c}, P4: \textit{b}) and a non-equilibrium strategy profile (P1: \textit{a}, P2: \textit{b}, P3: \textit{a}, P4: \textit{a}) lead to a total payoff of 66 (see Figure \ref{fig:ring_games_EQ}).} 
In this study, we employ an alternative approach that overcomes this limitation to measure rationality levels: letting a subject play against equilibrium-type computer players (i.e., the Robot Treatment).

Similar to the motivation of our Robot Treatment, \cite{devetag2003games}, \cite{grehl2015experimental}, and \cite{bayer2016logical} also employ rational computer players to mitigate the impact of beliefs and social preferences on individual decisions in their experiments.
While \cite{devetag2003games} find a positive correlation between short-term memory performance and conformity to standard theoretical predictions in strategic behavior, \cite{grehl2015experimental} and \cite{bayer2016logical} explore players' ability to reason logically about others' types in the incomplete information game known as the dirty faces game.
In contrast, our study departs from theirs by focusing on investigating whether playing against computers can provide a robust measure of strategic reasoning ability across different families of games with complete information.
Additionally, we also include a memory task to investigate whether the lack of significant predictive power of short-term memory on reasoning levels observed in GHW is influenced by uncontrolled beliefs and to offer a robustness check for the findings of \cite{devetag2003games} in different settings.

In previous studies on strategic reasoning, equilibrium-type computer players have been introduced into laboratory experiments to induce human players' equilibrium behavior (e.g., \citealp{costa2006cognition}; \citealp{meijering2012eye}) and to eliminate strategic uncertainty (e.g., \citealp{hanaki2016cognitive}).\footnote{For a survey of economics experiments with computer players, see \cite{march2021strategic}.} 
In contrast, our aim is to utilize computer players to uncover individual strategic reasoning ability. 
Our study contributes to the literature by demonstrating that introducing robot players can induce human subjects to exhibit stable reasoning levels across games, thus providing a solid foundation for measuring individual strategic ability.

\section{Theoretical Framework}\label{sec:theory}

\subsection{The Model in GHW}\label{sec:model_in_georganas}

To formalize the idea of the depth of rationality and the hypotheses we are going to test, we introduce the model and notations used in GHW. In their model, an $n$-person normal form game $\gamma \in \Gamma$ is represented by $(N, S, \{u_i\}_{i \in N})$, where $N = \{1,...,n\}$ denotes the set of players, $S = S_1 \times \cdots \times S_n = \Pi_{i = 1}^n S_i$ denotes the strategy sets, and $u_i : S \to \mathrm{R}$ for $i \in N$ denotes the payoff functions. Following GHW, we use $u_i(\sigma)$ to refer to $E_\sigma [u_i(\sigma)]$, where $\sigma = (\sigma_1,...,\sigma_n)$, when $\sigma$ is a profile of mixed strategies (i.e., $\sigma_i \in \Delta(S_i)$).

Player $i$'s strategic ability is modeled by two functions $(c_i, k_i)$. Let $T$ be the set of \textit{environmental parameters}, which captures the information a player observes about their opponents' cognitive abilities. The function $c_i : \Gamma \rightarrow \mathbb{N}_0$ represents $i$'s \textit{capacity} for game $\gamma$, and the function $k_i : \Gamma \times T \rightarrow \mathbb{N}_0$ represents $i$'s (realized) \textit{level} for game $\gamma$. A player's level for a game is bounded by their capacity, so $k_i(\gamma, \tau_i) \leq c_i(\gamma)$ for all $\gamma$, $\tau_i \in T$, and $i \in N$. The goal of our experiment is to measure $c_i(\gamma)$ and to test if $c_i(\gamma)$ (or $k_i(\gamma, \tau_i)$, after controlling for $\tau_i$) exhibits any stability across different games (see Section \ref{subsec:hypotheses} for further discussion).

\subsection{Level-\textit{k} Model and Higher-Order Rationality}

In GHW, a player's behavior is characterized by the standard level-$k$ model. 
Specifically, let $\nu : \N_0 \rightarrow \Delta(\N_0)$ be a player's belief about their opponents' levels. 
In a standard level-$k$ model, $\nu(m) = \mathds{1}\{m - 1\}$ for all $m \geq 1$, and a level-0 player $i$'s strategy is exogenously given as $\sigma_i^0 \in \Delta(S_i)$.  
A level-$k$ ($k \geq 1$) player $i$'s strategy ($\sigma_i^k$) is defined inductively as a best response to $\nu(k)$. 
Formally, for all $s'_i \in S_i$, $\sigma_i^k$ satisfies $u_i(\sigma_i^k, \sigma_{-i}^{\nu(k)}) \geq u_i(s'_i, \sigma_{-i}^{\nu(k)})$ where $\sigma_{-i}^{v(k)} = (\sigma_1^{k - 1},...,\sigma_{i - 1}^{k - 1}, \sigma_{i + 1}^{k - 1},..., \sigma_n^{k - 1})$. 
Notice that in order to pin down a level-$k$ player's strategy, we need to impose an assumption on the level-0 strategy. However, some studies have reported variations in level-0 actions and level-0 beliefs across individuals \citep{burchardi2014out, chen2018window}. 
Thus, an individual's identified level of reasoning can be sensitive to the structural assumptions under a level-$k$ model.

To avoid the ad hoc assumptions on level-0 players , we can instead define $k$th-order rationality \citep{bernheim1984rationalizable, pearce1984rationalizable, lim2016identifying} in the following way. 
Let $R_i^k(\gamma)$ be the set of strategies that survive $k$ rounds of iterated elimination of strictly dominated strategies (IEDS) for player $i$. 
In other words, a strategy $s_i$ is in $R_i^1(\gamma)$ if $s_i$ is a best response to some arbitrary $s_{-i}$, and $s_i$ is in $R_i^{k'}(\gamma)$ if $s_i$ is a best response to some $s_{-i} \in R_{-i}^{{k'} - 1}(\gamma)$ for $k' > 1$. 
We say that a player $i$ exhibits \textit{$k$th-order rationality} in $\gamma$ if and only if $i$ always plays a strategy in $R_i^k(\gamma)$. 
Equivalently, an individual exhibits $k$th-order rationality if and only if there is a $\sigma_{-i}^0$ such that the individual can be classified as a level-$k$ player in a standard level-$k$ model. 
Note that given any game $\gamma \in \Gamma$, $R_i^{k+1}(\gamma) \subset R_i^k(\gamma)$ for all $k \in \N_0$. In other words, a player exhibiting $k$th-order rationality also exhibits $j$th-order rationality for all $j \leq k$.

\section{The Games}\label{sec:games}

We study two classes of games: the four-player ring games used in \cite{kneeland2015identifying} for identifying individuals' higher-order rationality and a variant of the two-person guessing games first studied by \cite{costa2006cognition} and used in GHW for identifying players' level-\textit{k} types.

\subsection{Ring Games}

\begin{figure}[htbp!]
    \centering
    \includegraphics[width=0.9\textwidth]{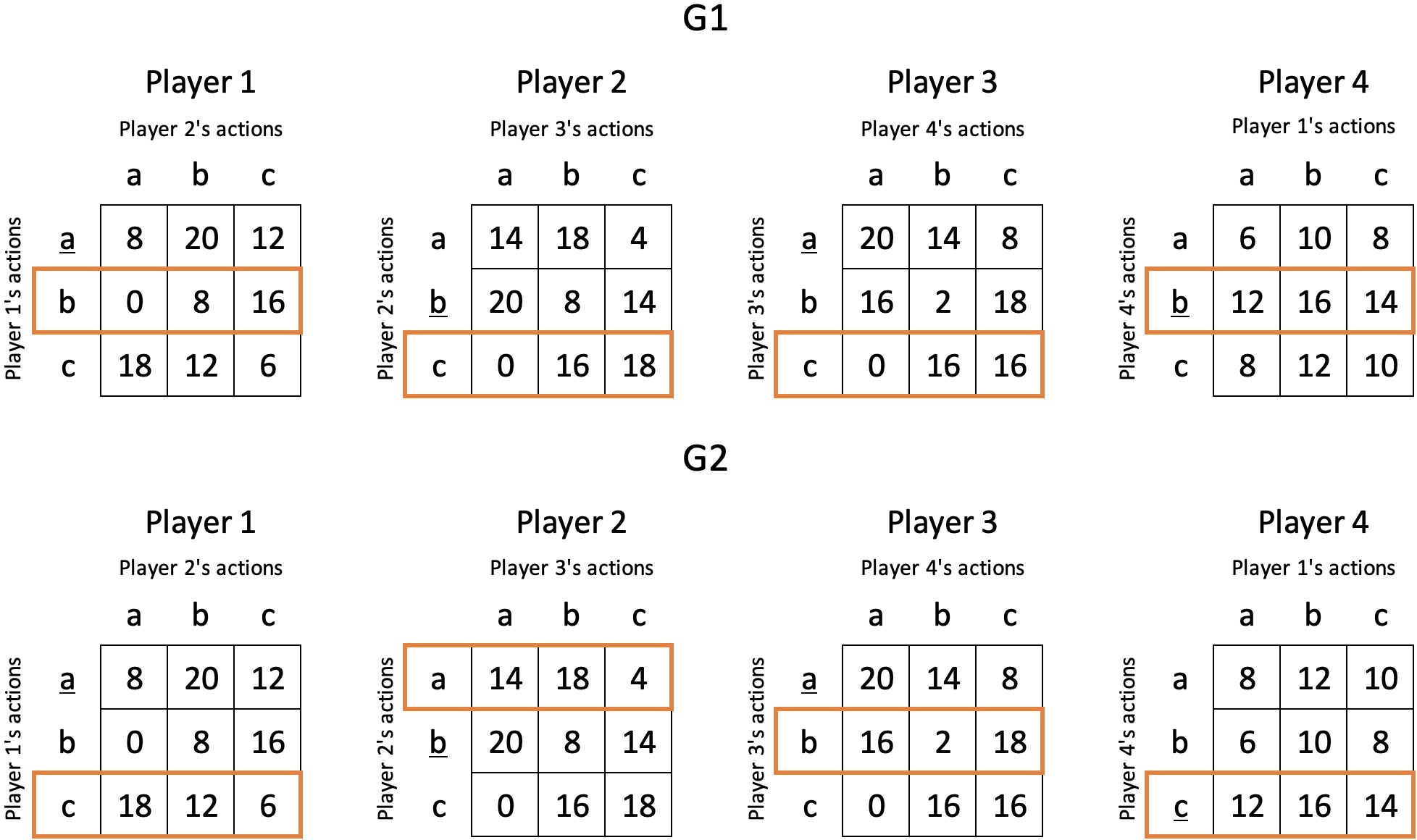}
    \caption{The Ring Games. The Nash Equilibrium is highlighted with colored borders, and the secure actions are underscored.}
    \label{fig:ring_games_EQ}
\end{figure}

A four-player ring game is a simultaneous game characterized by four $3 \times 3$ payoff matrices. Figure \ref{fig:ring_games_EQ} summarizes the structures of the two ring games, G1 and G2, used in our experiment. As shown in Figure \ref{fig:ring_games_EQ}, each player $i \in \{1, 2, 3, 4\}$ simultaneously chooses an action $a_i \in \{a, b, c\}$. Player 4 and Player 1's choices determine Player 4's payoff, and Player \textit{k} and Player (\textit{k} + 1)'s choices determine Player \textit{k}'s payoff for $k \in \{1, 2, 3\}$. 

The payoff matrices for Player 1, 2, and 3 are identical in G1 and G2. 
However, the matrix for Player 4 differs between G1 and G2, with the rows corresponding to Player 4's actions ($a$, $b$, $c$) interchanged, leading to different best replies in the subsequent matrices.

Specifically, Player 4 has a strictly dominant strategy in each ring game: $b$ in G1 and $c$ in G2. Given the payoff structure, a (first-order) rational individual will always choose \textit{b} in G1 and \textit{c} in G2 when acting as Player 4. By eliminating dominated strategies, an individual exhibiting second-order rationality will always choose $c$ in G1 and $b$ in G2 when acting as Player 3. Continuing this process iteratively, an individual exhibiting third-order rationality will always choose $c$ in G1 and $a$ in G2 when acting as Player 2, and an individual exhibiting fourth-order rationality will always choose $b$ in G1 and $c$ in G2 when acting as Player 1. Thus, the unique Nash equilibrium of G1 is Player 1, 2, 3, and 4 choosing $b$, $c$, $c$, and $b$, respectively, and for G2, Player 1, 2, 3, and 4 choosing $c$, $a$, $b$, and $c$, as highlighted in Figure \ref{fig:ring_games_EQ}.

Note that the payoff structures in our ring games are identical to those in \cite{kneeland2015identifying}, except that rows $a$ and $b$ are swapped for Player 4 in G1. 
This modification ensures that our equilibrium-predicted actions do not coincide with the \emph{secure} actions (or max-min actions) in both G1 and G2, which maximize the total payoff sum over the opponents' possible actions, potentially encouraging subjects to choose the equilibrium strategy based on non-payoff-maximizing motives.\footnote{A consequence of this modification is that the minimum possible payoff for the equilibrium strategy in G1 becomes 0 for Player 1, 2, and 3.}
Adopting the same payoff structure as Kneeland's design facilitates comparability between our results and hers.

\subsection{Guessing Games}

In our experiment, the guessing game is a simultaneous two-player game parameterized by a constant $p \in (0, 1)$. We use $p = 1/3$, $1/2$ and $2/3$ in our experiment. Each player $i$ simultaneously chooses a positive integer $s_i$ between 1 and 100. Player $i$'s payoff strictly decreases in the difference between the number chosen by $i$, $s_i$, and the number chosen by $i$'s opponent multiplied by a constant $p$, $p s_{-i}$. Specifically, player $i$'s payoff is equal to $0.2 \times (100 - |s_i - p s_{-i}|)$. Thus, a payoff-maximizing player's objective is to make a guess that matches their opponent's guess times $p$. 
Note that, given $p < 1$, any action (integer) greater than or equal to $\lfloor 100 p + 0.5 \rfloor + 1$ is strictly dominated by 
$ \lfloor 100 p + 0.5\rfloor $ 
since $|\lfloor 100 p + 0.5\rfloor  - p s_{-i}| < |s'_i - p s_{-i}|$ for all $s_{-i} \in \{1,...,100\}$ and 
$s'_i \in \{ \lfloor 100 p + 0.5\rfloor + 1,...,100\}$.\footnote{For instance, in a guessing game with $p = 1/3$, every integer between 34 and 100 is dominated by 33; when $p = 1/2$, every integer between 51 and 100 is dominated by 50; when $p = 2/3$, every integer between 68 and 100 is dominated by 67.}

Given the payoff function, a rational individual will always choose an integer between 1 and $K_1 \equiv \lfloor 100 p + 0.5\rfloor $. 
A second-order rational individual will believe the other player is first-order rational and choose 
a positive integer between 1 and $\lfloor K_1 p + 0.5\rfloor $, and so on. The unique equilibrium of the two-person guessing game is thus both players choosing 1.

\section{Identification}\label{sec:identification}

Our model does not allow us to directly identify one's higher-order rationality from choice data. For example, an equilibrium player will choose 1 in the guessing game with $p = 1/2$, while a player choosing 1 may have only performed one step of reasoning if their first-order belief is that their opponent guesses 2. Thus, observing a player $i$ choosing a strategy in $R_i^k(\cdot)$ for $k > 1$ (in a finite number of rounds) does not imply that $i$ exhibits $k$th-order rationality, which renders an individual's higher-order rationality unidentifiable. In fact, following the definition of $R_i^k(\cdot)$, we have $R_i^{k+1}(\cdot) \subset R_i^k(\cdot)$ for all $k \in \N_0$. Namely, every strategy (except for the dominated actions) can be rationalized by some first-order belief.

Following the rationale of higher-order rationality, we use the \textit{revealed rationality approach} (\citealp{lim2016identifying}; \citealp{brandenburger2019identification}; \citealp{cerigioni2019higher}) as our identification strategy. As explained below, this approach allows us to identify individual higher-order rationality in a dominance-solvable game. Under the revealed rationality approach, we say that a player $i$ exhibits $k$th-order revealed rationality if (and only if) we observe the player actually playing a strategy that can survive $k$ rounds of IEDS, i.e., $s_i \in R_i^k(\cdot)$. A subject is then identified as a $k$th-order (revealed-)rational player when they exhibit $m$th-order revealed rationality for $m = k$ but not for $m = k + 1$. That is, a player is classified into the upper bound of their (revealed) rationality level.\footnote{\cite{kneeland2015identifying} uses the \textit{exclusion restriction} (ER) as its identification strategy, assuming that a player with low order rationality does not respond to changes in payoff matrices positioned away from herself. However, \cite{lim2016identifying} show that more than three-quarters of their experimental subjects change their actions in two identical ring games, which suggests the failure of the ER assumption since a rational player is predicted to take the same action in two identical games under the exclusion restriction. Also, the ER assumption does not facilitate the identification of higher-order rationality in the guessing games since we cannot separate out first-order payoffs from higher-order ones.}

The idea behind the revealed rationality approach is the ``as-if'' argument: a subject $i$ selecting $s_i \in R_i^k(\cdot) \setminus R_i^{k+1}(\cdot)$ in finite observations behaves like a $k$th-order rational player, who always selects a strategy in $R_i^{k}(\cdot)$ but probably not in $R_i^{k+1}(\cdot)$, and thus is identified as a $k$th-order revealed rational player. Under this identification criterion, we can identify an individual's order of (revealed) rationality without requiring them to play in multiple games with different payoff structures. In our data analysis, we will classify subjects into five different types: first-order revealed rational (R1), second-order revealed rational (R2), third-order revealed rational (R3), fourth-order (or fully) revealed rational (R4), and non-rational (R0).\footnote{In a four-player ring game, the highest identifiable (revealed) order of rationality is level four.} Tables \ref{tab:reveal_actions_ring} and \ref{tab:reveal_actions_guessing} summarize the predicted actions under the revealed rationality approach for each type of players in our ring games and guessing games, respectively.

\begin{table}[htbp!]
    \centering
    \caption{Predicted Actions in the Ring Games Under the Revealed Rationality Approach}
    \label{tab:reveal_actions_ring}
    \begin{tblr}{
      column{1} = {c},
      column{3} = {c},
      column{4} = {c},
      column{6} = {c},
      column{7} = {c},
      column{9} = {c},
      column{10} = {c},
      column{12} = {c},
      column{13} = {c},
      cell{1}{3} = {c=11}{c},
      cell{2}{3} = {c=2}{c},
      cell{2}{6} = {c=2}{c},
      cell{2}{9} = {c=2}{c},
      cell{2}{12} = {c=2}{c},
      cell{4}{3} = {c=2}{c},
      cell{4}{6} = {c=2}{c},
      cell{4}{9} = {c=2}{c},
      cell{4}{12} = {c=2}{c},
      cell{5}{3} = {c=2}{c},
      cell{5}{6} = {c=2}{c},
      cell{5}{9} = {c=2}{c},
      cell{5}{12} = {c=2}{c},
      cell{6}{3} = {c=2}{c},
      cell{6}{6} = {c=2}{c},
      cell{6}{9} = {c=2}{c},
      cell{6}{12} = {c=2}{c},
      cell{7}{3} = {c=2}{c},
      cell{7}{6} = {c=2}{c},
      cell{7}{9} = {c=2}{c},
      cell{7}{12} = {c=2}{c},
      cell{8}{3} = {c=2}{c},
      cell{8}{6} = {c=2}{c},
      cell{8}{9} = {c=2}{c},
      cell{8}{12} = {c=2}{c},
      hline{1,4,9} = {-}{},
      hline{2} = {3-13}{},
      hline{3} = {3-4,6-7,9-10,12-13}{},
    }
         &  & Ring Games &    &  &            &    &  &            &    &  &            &    \\
         &  & P1         &    &  & P2         &    &  & P3         &    &  & P4         &    \\
    Level &  & G1         & G2 &  & G1         & G2 &  & G1         & G2 &  & G1         & G2 \\
    R0   &  & N/A        &    &  & N/A        &    &  & N/A        &    &  & not (b, c) \phantom{not} &    \\
    R1   &  & N/A        &    &  & N/A        &    &  & not (c, b) \phantom{not} &    &  & (b, c)     &    \\
    R2   &  & N/A        &    &  & not (c, a) \phantom{not} &    &  & (c, b)     &    &  & (b, c)     &    \\
    R3   &  & not (b, c) \phantom{not} &    &  & (c, a)     &    &  & (c, b)     &    &  & (b, c)     &    \\
    R4   &  & (b, c)     &    &  & (c, a)     &    &  & (c, b)     &    &  & (b, c)     &    
    \end{tblr}
\end{table}

\begin{table}[htbp!]
\centering
\caption{Predicted Actions in the Guessing Games Under the Revealed Rationality Approach}
\label{tab:reveal_actions_guessing}
\begin{tblr}{
  column{odd} = {r},
  column{1} = {c},
  cell{1}{3} = {c=5}{c},
  cell{2}{3} = {c},
  cell{2}{5} = {c},
  cell{2}{7} = {c},
  cell{3}{4} = {r},
  cell{3}{6} = {r},
  cell{4}{4} = {r},
  cell{4}{6} = {r},
  cell{5}{4} = {r},
  cell{5}{6} = {r},
  cell{6}{4} = {r},
  cell{6}{6} = {r},
  cell{7}{4} = {r},
  cell{7}{6} = {r},
  hline{1,8} = {-}{},
  hline{2} = {3-7}{},
  hline{3} = {1,3,5,7}{},
}
              &  & Guessing Games &  &           &  &           \\
Level          &  & $p$ = 1/3      &  & $p$ = 1/2 &  & $p$ = 2/3 \\
R0            &  & 34--100         &  & 51--100    &  & 68--100    \\
R1            &  & 12--33          &  & 26--50     &  & 46--67     \\
R2            &  & 5--11           &  & 14--25     &  & 31--45     \\
R3            &  & 2--4            &  & 8--13      &  & 21--30     \\
R4 (or above) &  & 1              &  & 1--7       &  & 1--20      
\end{tblr}
\end{table}

\section{Experimental Design and Hypotheses}\label{sec:experiment_design}
\subsection{Treatments}
\label{subsec:treatments}

We design a laboratory experiment to measure subjects' higher-order rationality. 
In the main part of the experiment, subjects first play the ring games, 
followed by the guessing games, in two different scenarios: the \emph{Robot Treatment} and 
the \emph{History Treatment}. Using a within-subject design, we alternate the order of the two scenarios (RH Order and HR Order) across sessions to balance out potential spillover effects from one treatment to another.

In each scenario, each subject first plays the two four-player, three-action ring games 
(G1 and G2 in Figure \ref{fig:ring_games_EQ}) in each position in each game once (for a total of eight rounds). 
Each subject is, in addition, assigned a neutral label (Member A, B, C, or D) before the ring games start. 
The label is only used for the explanation of an opponent's strategy in the History Treatment and does not reflect player position.
To facilitate the cross-subject comparison, all the subjects play the games in the following 
fixed order: P1 in G1, P2 in G1, P3 in G1, P4 in G1, P1 in G2, P2 in G2, 
P3 in G2, and P4 in G2.\footnote{Note that Player 4 has a dominant strategy in the ring game. 
We have our subjects play in each position in the reverse order of the IEDS procedure to
mitigate potential framing effects resulting from the hierarchical structure.}
The order of payoff matrices is also fixed, with a subject's own payoff matrix being fixed 
at the leftmost side.\footnote{This feature is adopted in \cite{jin2021does} and the main
treatment of \cite{kneeland2015identifying}. \cite{kneeland2015identifying} perturbs the order 
of payoff matrices in a robust treatment and finds no significant effects on subject behavior.}

In the Robot Treatment, the subjects play against fully rational computer players. 
Specifically, each subject in each round is matched with three robot players who only 
select the strategies that survive iterated dominance elimination (i.e., the equilibrium strategy). 
We inform the subjects of the presence of robot players that exhibit third-order rationality.\footnote{Since level four is the highest 
identifiable (revealed) order of rationality in a four-player ring game, 
incorporating a third-order rational computer player is sufficient
to identify this maximum level.} 
The instructions for the robot strategy are as 
follows:\footnote{Our instructions are adapted from the experiment 
instructions of Study 2 of \cite{johnson2002detecting}. The original instructions
are as follows: ``In generating your offers, or deciding whether to accept or reject offers, 
assume the following: 1. You will be playing against a computer which is 
programmed to make as much money as possible for itself in each session. The 
computer does not care how much money you make. 2. The computer program expects you to try 
to make as much money as you can, and the program realizes that you have been told, in 
instruction (1) above, that it is trying to earn as much money as possible for itself'' (p.\@ 44-45).} 
\begin{quote}
    \emph{When you start each new round, you will be grouped with three other participants who are in different roles. The three other participants will be computers that are programmed to take the following strategy:
\begin{enumerate}
    \item The computers aim to earn as much payoff as possible for themselves.
    \item A computer believes that every participant will try to earn as much payoff as one can.
    \item A computer believes that every participant believes ``the computers aim to earn as much payoff as possible for themselves.''
\end{enumerate}
}
\end{quote}
The first line of a robot's decision rule (``The computers aim to...'') implies that a robot 
never plays strictly dominated strategies and thus exhibits first-order rationality. 
The second line (along with the first line) indicates that a robot holds the belief that
other players are (first-order) rational and best responds to such belief, which 
implies a robot's second-order rationality. The third line (along with the first and second lines)
implies that, applying the same logic, a robot exhibits third-order rationality.

In the History Treatment, the subjects play against the data drawn from their decisions in the previous scenario.
Specifically, in each round, a subject is matched with three programmed players who adopt
actions chosen in the Robot Treatment by three other subjects.\footnote{In the HR Order sessions, 
the choices made by a subject's opponents were drawn from the participants in the Robot Treatment of previous sessions.} 
Every subject is informed that other human participants' payoffs would not be affected by their choices at this stage.
By having the subjects play against past decision data, we can exclude
the potential confounding effect of other-regarding preferences on individual actions.

After the ring games, the subjects play the two-person guessing games (in the order
of $p = 2/3$, $1/3$, $1/2$) in both the Robot Treatment and the History Treatment. 
Instead of being matched with three opponents, a subject is matched with only one player in the guessing games. 
The instructions for the guessing games in both treatments are revised accordingly.

\subsection{Hypotheses}
\label{subsec:hypotheses}

The Robot Treatment is designed to convince subjects that the computer opponents 
they face are the most sophisticated players they could encounter.
Consequently, if our Robot Treatment is effectively implemented, it should prompt subjects to employ a strategy at the highest achievable level $k$, i.e., $k_i(\gamma, \tau_i = Robot) = c_i(\gamma)$ for all $\gamma$ and $i$. 
(Recall that $k_i$ and $c_i$ denote subject $i$'s realized level and capacity, respectively.) 
This observation gives rise to the first hypothesis we aim to evaluate.

\bigskip
\noindent\textbf{Hypothesis 1 (Bounded Capacity).} 
$k_i(\gamma, \tau_i = History) \leq k_i(\gamma, \tau_i = Robot)$ for all $\gamma$.
\bigskip

In other words, we test whether subjects' rationality levels against robots capture individual strategic reasoning capacity.
The corresponding analysis is presented in Section \ref{subsection:result_type}.

If Hypothesis 1 holds, then we can evaluate several possible restrictions on $c_i$ by forming hypotheses on $k_i(\gamma, Robot)$. In evaluating Hypothesis 2, we examine whether there are stable patterns in (revealed) individual reasoning depth across games. 

\bigskip
\noindent\textbf{Hypothesis 2 (Constant Capacity).} 
$k_i(\gamma, Robot) = k_i(\gamma', Robot)$ for all $\gamma, \gamma'$.
\bigskip

This hypothesis imposes the strictest requirement on stability by testing if a player's rationality level remains constant across games. 
In other words, it assesses whether playing against robots provides a measure of one's \textit{absolute} depth of reasoning. 
The corresponding analysis is presented in Section \ref{subsection:result_constant}.

In addition to these two hypotheses, \ref{appendix:additional_results} explores 
two less stringent stability requirements, such as the stability of relative rankings 
between players' rationality levels (Hypothesis 3) or the consistency of
game difficulty in terms of revealed rationality across players 
(Hypothesis 4).\footnote{Our Hypothesis 2, 3, and 4 correspond to Restriction 2, 3, and 5 in GHW, respectively (see p.\@ 377).}

\subsection{Cognitive Tests}
\label{subsec:cognitive_tests}

Apart from the ring games and the guessing games, subjects also complete
three cognitive tests to measure different aspects of their cognitive ability and strategic reasoning:
\begin{itemize}
    \item[(1)] the Cognitive Reflection Test (CRT), 
    \item[(2)] the Wechsler Digit Span Test, and 
    \item[(3)] the farsightedness task.
\end{itemize}

The CRT, proposed by \cite{frederick2005cognitive}, is designed to evaluate the ability to reflect on intuitive answers.
This test contains three questions that often 
trigger intuitive but incorrect answers. Performance on this test has been found to be 
correlated with strategic abilities. For instance, 
GHW report that the subjects' CRT scores have moderate predictive power on their expected earnings and level-\textit{k} types.

The second test is the Wechsler Digit Span Test \citep{wechsler1939measurement}, 
which is designed to test short-term memory. In our experiment, this test contains eleven rounds. 
In each round, a subject needs to repeat a sequence of digits displayed on the screen
at the rate of one digit every second. The maximum length of the digit sequence a subject
can memorize reflects the subject's short-term memory 
capacity.\footnote{The length of the digit sequence increases from three digits to thirteen digits round by round.} \cite{devetag2003games} find a positive correlation between individual short-term memory and strategic ability.

\begin{figure}[htbp!]
    \centering
    \includegraphics[width=.5\textwidth]{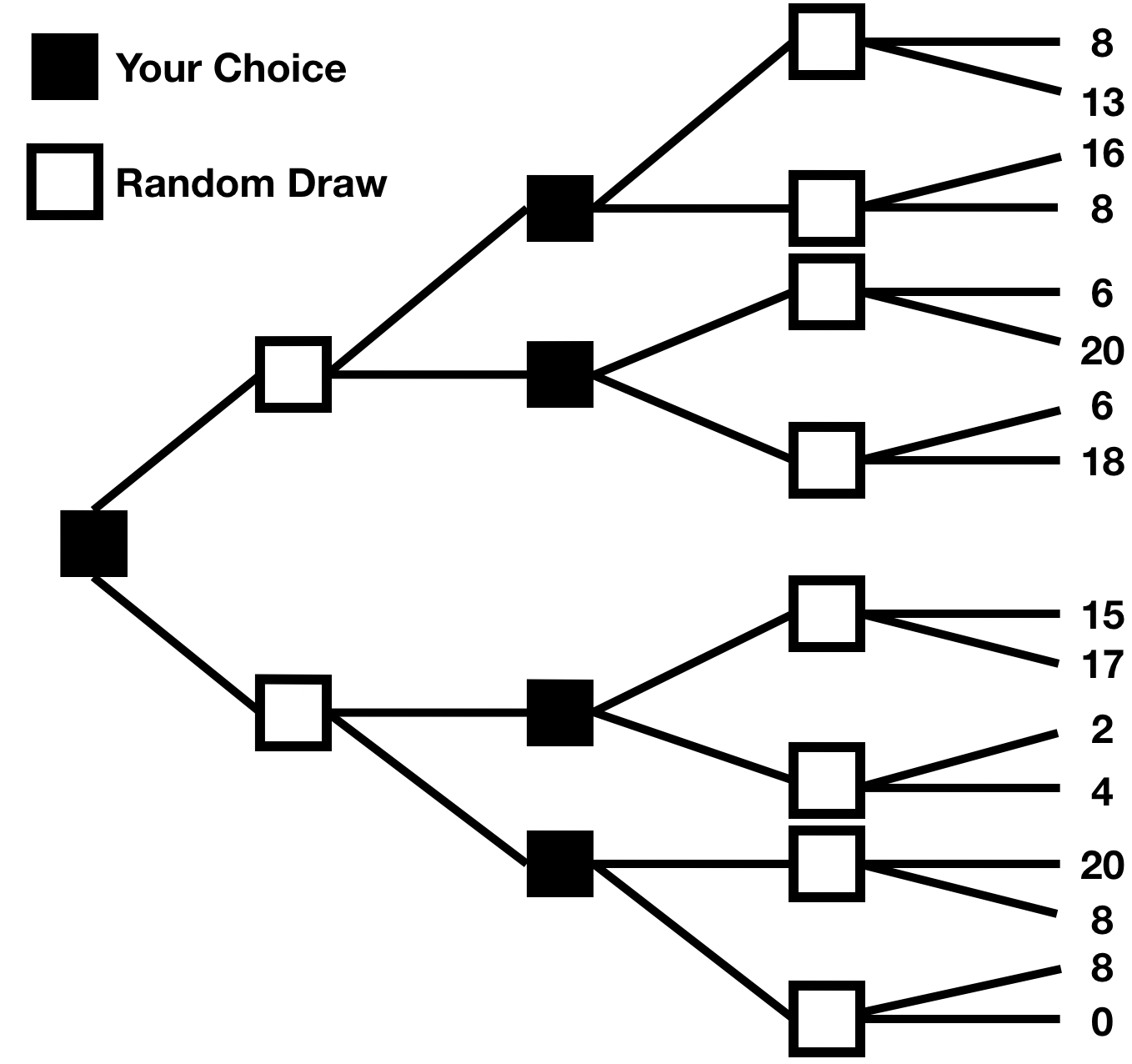}
    \caption{The Farsightedness Task in \cite{bone2009people}}
    \label{fig:farsight}
\end{figure}

Lastly, the \emph{farsightedness task}, developed by \cite{bone2009people}, is an 
individual task to measure a subject's ability to do backward induction, 
or to anticipate their own future action and make the best choice accordingly.
{Specifically, it} is a sequential task that involves two sets of
decision nodes and two sets of chance nodes (see the decision tree 
in Figure \ref{fig:farsight}). The first and third sets of
nodes are the decision nodes where a decision maker is going to take an
action (up or down). The second and fourth sets of nodes are the chance
nodes where the decision maker is going to be randomly assigned an action (with equal probability). 

Notice that there is one dominant action, in the sense of first-order stochastic dominance, 
at each of the third set of nodes (i.e., the second set of decision nodes). 
Anticipating the dominant actions at the second set of decision nodes, 
the decision maker also has a dominant action (down) at the first node. However, 
if a payoff maximizer lacks farsightedness and anticipates that each payoff will be
reached with equal chance, then the dominated action (up) at the first node will become 
the dominant option from this decision maker's perspective. 
Therefore, a farsighted payoff-maximizer is expected to choose down, but a myopic one
is expected to choose up, at the first move (and choose the dominant actions at the second moves).
Consequently, we can use their choice at the first move to evaluate
the correlation between one's farsightedness and rationality level.

\subsection{Laboratory Implementation}

We conducted 41 sessions between August 31, 2020 and January 28, 2021 at the 
Taiwan Social Sciences Experiment Laboratory (TASSEL) in National Taiwan University (NTU). 
The experiment was programmed with the software zTree \citep{fischbacher2007z} and instructed in Chinese.
A total of 299 NTU students participated in the experiment, all recruited 
through ORSEE \citep{greiner2015subject}.
In our experiment, 136 subjects played the Robot Treatment before the History Treatment in both families of
games (RH Order), while 157 subjects played the History Treatment first 
(HR Order).\footnote{Six subjects are dropped from our analysis due to computer crashes.}

\begin{figure}[htbp!]
    \centering
    \includegraphics[width=\textwidth]{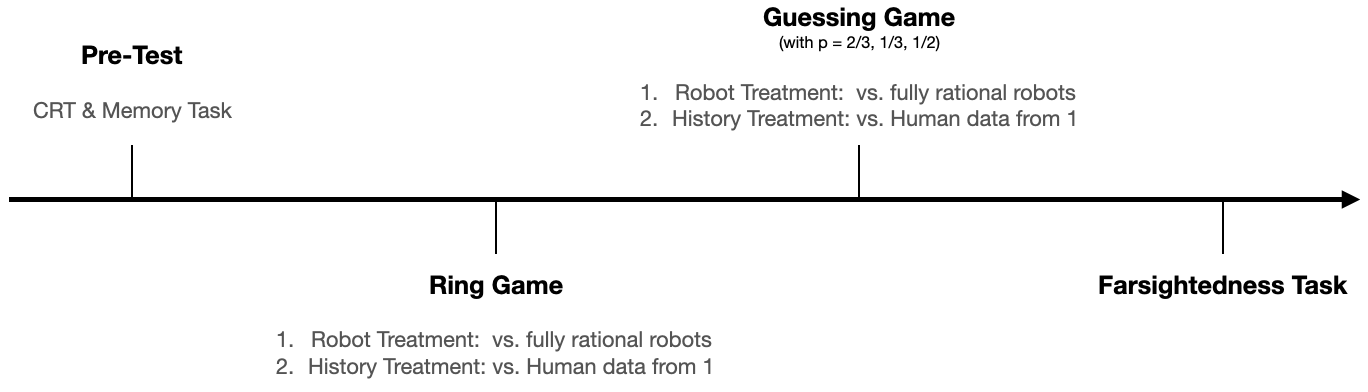}
    \caption{Experiment Protocol}
    \label{fig:protocol}
\end{figure} 

Each experimental session lasted about 140 minutes, and 
the protocol is summarized in Figure \ref{fig:protocol}.
At the beginning of the experiment, subjects first completed 
the CRT and the Wechsler Digit Span Test. After these tasks, 
subjects played the ring games in both the Robot Treatment and the History Treatment, 
followed by the guessing games in both treatments. In the final section of the experiment, 
subjects were asked to complete the farsightedness task.
The experimental subjects did not receive any feedback about the outcomes of their choices until the end of the experiment.

There was a 180-second time limit on every subject's decisions in the ring games, guessing games, and farsightedness task. 
A subject who did not confirm their choice within 180 seconds would have 
earned zero payoff for that round; however, no subjects 
exceeded the time limit.\footnote{\cite{jin2021does} sets a 60-second time limit on 
decisions in the ring games and finds little effect on type classification.}

The subjects were paid based on the payoffs (in ESC, Experimental Standard Currency) 
they received throughout the experiment. In addition to the payoff in the 
farsightedness task, one round in the ring games and one round in the guessing 
games were randomly chosen for payment. 
A subject also got three ESC for each correct answer in the CRT, 
and one ESC for each correct answer in the Digit Span Test. 
Including a show-up fee of NT\$200 (approximately \$7 in USD in 2020), 
the earnings in the experiment ranged between NT\$303 and NT\$554, with an average 
of NT\$430.\footnote{The exchange rate was 1 ESC for NT\$4, and the foreign exchange rate was around US\$1 $=$ NT\$29.4.}

\section{Experiment Results}\label{sec:results}

In this section, we first provide a general 
description of the data in Section
\ref{subsection:result_data_description}. Next, 
we classify subjects into different rationality levels 
using the revealed rationality approach in Section 
\ref{subsection:result_type}, showing that subjects display higher levels of rationality when playing against robots.
In Section \ref{subsection:result_constant}, we
demonstrate that individual rationality levels are significantly 
more stable when controlling for subjects' beliefs about 
their opponents' depth of reasoning.
Finally, we explore the correlation between depth of 
reasoning, performance on cognitive tests and the 
heuristics of choosing secure actions in 
Section \ref{subsection:result_correlation}.

\subsection{Data Description}
\label{subsection:result_data_description}

\begin{figure}[htbp!]
    \centering
    \includegraphics[width=0.88\textwidth]{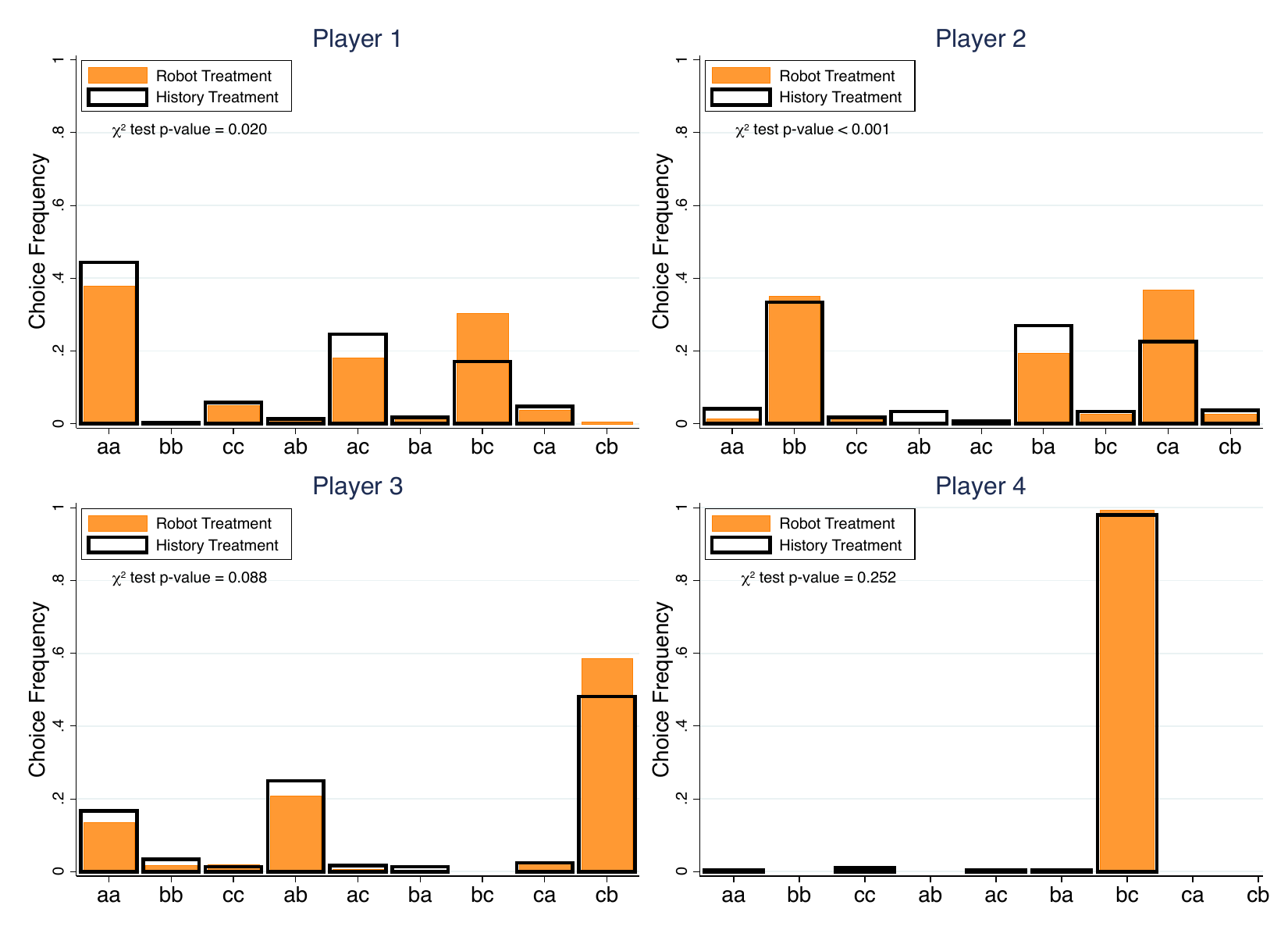}
    \caption{Ring Game Choice Frequency at Each Position. The first and the second arguments represent the actions of G1 and G2.}
    \label{fig:pool_ring_choices}
\end{figure}

\begin{figure}[htbp!]
    \centering
    \includegraphics[width=\textwidth]{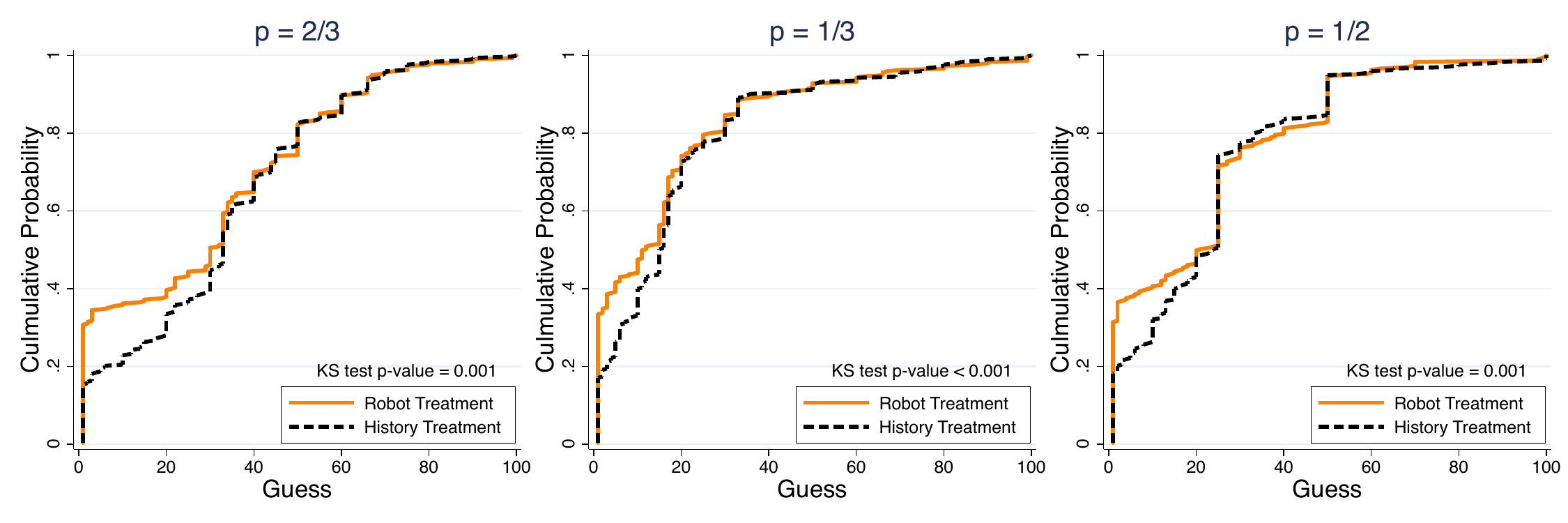}
    \caption{Cumulative Distribution of Guesses}
    \label{fig:pool_guess_choices}
\end{figure}

Before delving into the main results, we begin by summarizing the subjects' choice frequencies in the ring games (Figure \ref{fig:pool_ring_choices}) and guessing games (Figure \ref{fig:pool_guess_choices}). 
Figure \ref{fig:pool_ring_choices} reports the subjects' choice frequencies in the two ring games (G1 and G2, see Figure \ref{fig:ring_games_EQ}) at each player position.
From the figure, we can first observe that in both treatments, over 97\% of subjects choose the equilibrium strategy ($b$, $c$)
at P4 ($\chi^2$ test $p$-value = 0.252). This suggests 
that subjects are able to recognize strict dominance in the ring games.

Second, at each player position except P4, the significance of $\chi^2$ tests suggests that subjects' behavior 
is responsive to the treatments (P1: $\chi^2$ test $p$-value = 0.020; P2: $\chi^2$ test $p$-value
$<0.001$; P3:  $\chi^2$ test $p$-value = 0.088). 
Moreover, the Robot Treatment shows a 10 to 15 percentage point higher frequency of subjects 
choosing the equilibrium strategy ($b$, $c$) at P1, ($c$, $a$) at P2, and ($c$, $b$) at P3 compared to the History Treatment, 
indicating that the Robot Treatment effectively prompts subjects to display higher rationality levels.

Third, at each player position except P4, a notable proportion of subjects choose the secure action that maximizes the minimum possible payoff among the three available actions ($a$ at P1, $b$ at P2, $a$ at P3).
As shown in Figure \ref{fig:pool_ring_choices}, 
a high proportion of subjects opt for secure actions as an alternative to equilibrium actions. Moreover, except at P4,
the proportion of secure actions is higher in earlier positions. At P1, 38\% of subjects in the Robot Treatment and 44\% in the History Treatment
choose the secure actions ($a$, $a$). This tendency is more 
pronounced in the History Treatment, where secure actions are chosen more frequently than equilibrium actions at P1 and P2.\footnote{In the Robot Treatment at P2, the secure action profile $(b, b)$ and the equilibrium action profile $(c, a)$ 
are chosen 35\% and 37\% of the time, respectively. By contrast, 
in the History Treatment, the secure action profile and the equilibrium action profile are chosen 33\% and 23\% of the time, respectively.}
This evidence suggests that when players have uncertainty about their opponents' reasoning and strategic behavior, some players may opt for a non-equilibrium strategy to avoid the possibility of 
experiencing the worst possible payoff.\footnote{It is also worth noting
that at P1 and P2, compared to the Robot Treatment, 
action profiles involving secure actions in G1 and equilibrium actions in G2 
(i.e., ($a$, $c$) at P1 and ($b$, $a$) at P2) are more frequently 
observed in the History Treatment. The empirical frequency of 
action profile ($a$, $c$) at P1 is 18\% in the Robot Treatment
but 25\% in the History Treatment. Similarly, the 
frequency of action profile ($b$, $a$) at P2 is 19\% in the 
Robot Treatment and 27\% in the History Treatment. One potential 
reason is that choosing ($a$, $c$) at P1 and ($b$, $a$) at P2 are the
empirical best response in the History Treatment, and this behavior could be 
highly rational under a more general notion of 
rationalizability \citep{germano2020uncertain}.  
See \ref{appendix:additional_results}
for the analysis of the empirical best response in the History Treatment.}
A detailed analysis of 
the behavior of choosing secure actions is provided in Section 
\ref{subsection:result_correlation}.

Figure \ref{fig:pool_guess_choices} 
presents the cumulative distribution of subjects' guesses across the three guessing games.
We observe significant
differences in the distributions between the two treatments, regardless of the value of $p$ ($p=2/3$: KS test 
$p$-value = 0.001; $p=1/3$: KS test $p$-value < 0.001; $p=1/2$: KS test $p$-value = 0.001). Furthermore, 
in the Robot Treatment, there is a 13 to 16 percentage point higher proportion of subjects making the equilibrium guess 
(i.e., choosing 1) across all three guessing games compared to the History Treatment.
This difference leads to
first order stochastic dominance of the cumulative distribution of guesses in the Robot Treatment over that in the History Treatment, indicating a higher rationality
level among subjects in the Robot Treatment for the guessing games.

Furthermore, these distributional differences are 
driven by variations in equilibrium choices (i.e., 1). After 
excluding the equilibrium choice of 1, the cumulative 
distributions between the two treatments are not significantly 
different for any value of $p$ ($p=2/3$: KS test $p$-value = 
0.218; $p=1/3$: KS test $p$-value = 0.704; $p=1/2$: KS test 
$p$-value = 0.129). This result further confirms that subjects prompted to perform at their maximum depth of reasoning when facing robots are the primary driving force behind the deeper reasoning observed in the Robot Treatment.
In the next section, we will describe our approach for classifying individual rationality levels and perform statistical tests to assess whether subjects demonstrate higher rationality levels when playing against robots.

\subsection{Rationality Level Classification}
\label{subsection:result_type}

We adopt the revealed rationality approach to classify subjects into different rationality levels. 
Specifically, let $s_i = (s^{\gamma}_i)$ be the vector which collects player $i$'s actions in each family of games $\gamma$, where $\gamma \in \{\mbox{Ring}, \mbox{Guessing}\}$. 
In the ring games, we classify subjects based on the classification rule shown in Table \ref{tab:reveal_actions_ring}.
In both the Robot Treatment and the History Treatment, if a subject's action profile matches one of the predicted action profiles of type R0--R4 exactly, then the subject is assigned that level.  
Therefore, we can obtain each subject's rationality level in the Robot Treatment and the History Treatment, which are denoted as $k_i(\mbox{Ring}, \mbox{Robot})$ and 
$k_i(\mbox{Ring}, \mbox{History})$, respectively.

Similarly, for the guessing games, we classify subjects based on the rule outlined in Table \ref{tab:reveal_actions_guessing}. 
In both treatments, each subject makes three guesses (at $p=2/3$, $1/3$, and $1/2$). 
If a subject is categorized into different levels in different guessing games, we assign the subject the lower level. 
Thus, we can obtain the levels in both treatments, denoted as $k_i(\mbox{Guessing}, \mbox{Robot})$ and $k_i(\mbox{Guessing}, \mbox{History})$, respectively. 
Following this rationale, we construct the overall distribution of individual rationality levels for each treatment by assigning each subject the lower level they exhibit across the two classes of games, i.e., $k_i(\tau_i) = \min\{k_i(\mbox{Ring}, \tau_i), k_i(\mbox{Guessing}, \tau_i)\}$.\footnote{An alternative method for estimating overall levels across games is to impose a probabilistic error structure on deviations from predicted actions (e.g., \citealp{stahl1994experimental, stahl1995players}). However, this is incompatible with the revealed rationality framework, which does not predict a unique best action for each type. Additionally, assigning subjects to the lower order provides a reserved estimate, allowing for a more conservative test when comparing types between the Robot and History Treatments, thus increasing confidence if a statistical difference is observed.}

\begin{figure}[htbp!]
    \centering
    \includegraphics[width=0.9\textwidth]{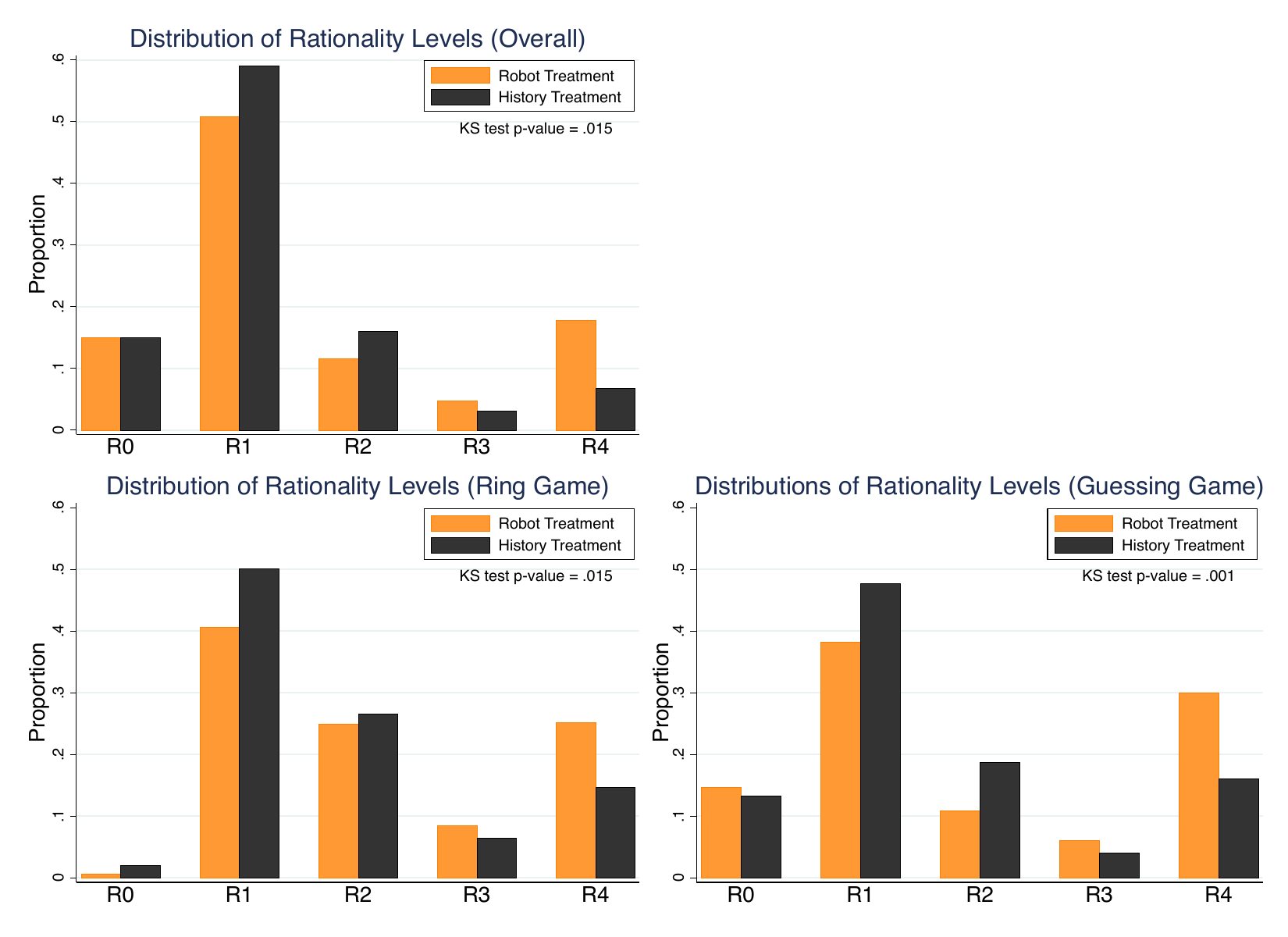}
    \caption{Distributions of Rationality Levels. The 
    top figure is the overall distribution of rationality 
    levels. The bottom figures are the distributions 
    of rationality levels in ring games (Left) and 
    guessing games (Right).} 
    \label{fig:level}
\end{figure}

Figure \ref{fig:level} reports the 
overall distribution of rationality levels for 
the Robot and History Treatments. As shown in the 
top figure, subjects tend to be classified into higher levels 
when playing against robots.
There are more R1 and R2 players but fewer R3 and R4 players in the History Treatment than in the Robot Treatment.
To examine if a subject's reasoning depth is bounded by their revealed rationality level in the Robot Treatment (Hypothesis 1), at the aggregate level, we conduct the two-sample Kolmogorov-Smirnov test to compare the distributions of rationality levels in the two treatments. 
If Hypothesis 1 holds, we should observe either no difference in the two distributions or the distribution in the Robot Treatment dominating the distribution in the History Treatment.
Our results show that the underlying distribution of individual rationality levels in the Robot Treatment stochastically dominates the one in the History Treatment 
(KS test $p$-value = 0.015), and thus provide supporting evidence for Hypothesis 1.

This result is robust across different types of games.
As shown in the bottom panels of 
Figure \ref{fig:level}, a similar pattern of 
first order stochastic dominance is observed regardless of whether rationality levels are 
classified based on behavior in the ring games or
the guessing games (Ring game: KS test $p$-value = 0.015; Guessing game: KS test $p$-value = 0.001).

Moreover, our within-subject design gives us paired data of individual rationality levels across treatments, which gives us another way to test Hypothesis 1. 
Overall, 85 percent of subjects (249/293) exhibit (weakly) higher rationality levels in the Robot Treatment than in the History Treatment. 
We further conduct the Wilcoxon signed-rank test to examine whether the subjects' rationality levels in the Robot Treatment are significantly greater than the History Treatment. 
Consistent with Hypothesis 1, we observe higher rationality levels in the Robot Treatment (Wilcoxon test $p$-value $ < 0.0001$). 
Therefore, we conclude that the rationality levels in the Robot Treatment can serve as a proxy of individual strategic reasoning capacity. 
In \ref{appendix:additional_results},
we separate the data by different games, 
finding a robust pattern in both.

It is noteworthy that, contrary to previous findings, we observe very few R0 players in the ring games in both treatments (Robot: 0.68\%; History: 2.04\%).\footnote{\cite{kneeland2015identifying} observes 6 percent of R0 players (with the ER approach) and \cite{cerigioni2019higher} observe more than 15 percent of R0 players (with the revealed rationality approach) in their experiments.} 
In our experiment, the subjects do not interact with each other in both treatments. 
Thus, our observation suggests that, when human interactions exist, social preferences may play some roles in a ring game and lead to (seemingly) irrational behavior, though we cannot exclude the possibility that this discrepancy in the prevalence of R0 players is due to different samples.

Yet in the guessing games, our classification results display a typical distribution pattern of estimated levels as documented in \cite{costa2006cognition} and GHW.
First, the modal type is R1 (Level 1), with more than 35 percent of subjects classified as R1 players in both treatments (Robot: 38.23\%; History: 47.78\%; \cite{costa2006cognition}: 48.86\%; GHW: 50.00\%).
In particular, the proportion of R1 players reported in the History treatment of our guessing games is very close to the proportion of level-1 players reported in \cite{costa2006cognition}  and GHW.
Second, R3 (Level 3) represents the least frequently observed category among the rational types (i.e., R1--R4), with fewer than 10 percent of subjects classified as R3 players in both treatments, a proportion that aligns with findings in the literature. (Robot: 6.14\%; History: 4.10\%; \cite{costa2006cognition}: 3.41\%; GHW: 10.34\%). 
Third, the percentage of R4 players in our History Treatment falls within the range of equilibrium-type player proportions reported in \cite{costa2006cognition}  and GHW (Robot: 30.03\%; History: 16.04\%; 
\cite{costa2006cognition}: 15.91\%; GHW: 27.59\%).
Noticeably, in our Robot Treatment, we observe a relatively high frequency of R4 players compared to previous literature.\footnote{For instance, \cite{arad201211} also note that, in their 11--20 money request game, the percentage of subjects employing more than three steps of iterative reasoning does not exceed 20 percent. This aligns with the proportion of R4 players identified in our History Treatment but is lower than that in our Robot Treatment.}
This finding underscores the significant impact of non-equilibrium belief about opponents on non-equilibrium behavior.

While our subjects' revealed rationality levels are comparatively higher when playing against robots, most do not exhibit more than two steps of reasoning.
In the Robot Treatment, around 70 percent of subjects still show an overall rationality level below the third order. 
This result supports the long-standing idea in the level-$k$ literature: humans have a relatively low cognitive ceiling for strategic thinking, often below level four.

\subsection{Consistency of Rationality Levels Across Games}
\label{subsection:result_constant}

In this section, we evaluate whether 
controlling for beliefs about the opponent's depth of reasoning 
leads individuals to reveal consistent rationality levels across 
games. There are different notions of consistency. As a first
exercise, we assess the strictest form of consistency: 
an individual reveals \emph{constant} rationality levels 
across games (Hypothesis 2). 

To examine this hypothesis, we generate a Markov transition matrix of rationality levels between the ring games and 
the guessing games in the Robot Treatment. 
Table \ref{tab:markov_robot}
reports the frequency with which an individual
moves from each rationality level in the ring games 
to each rationality level in the guessing games in the 
Robot Treatment.
If the observed individual rationality level is the same across games, then every diagonal entry of each transition matrix in Table \ref{tab:markov_robot} will be 100\%. 
Alternatively, if subjects' rationality levels in the ring games and guessing games are uncorrelated, every row in a transition matrix will be the same and equals the overall distribution in the guessing games.

\begin{table}[htbp!]
\centering
\begin{threeparttable}
\caption{Markov Transition for Rationality Levels
in the Robot Treatment}
\label{tab:markov_robot}
\renewcommand{\arraystretch}{1.3}
\begin{tabular}{rcccccc}
\hline
\multicolumn{1}{c}{} &  & \multicolumn{5}{c}{Guessing Games} \\
\multicolumn{1}{l}{From $\downarrow$ to $\rightarrow$} &  & R0 & R1 & R2 & R3 & R4 \\ \hline
\multicolumn{1}{l}{Ring Games} &  &  &  &  &  &  \\
R0 &  & \cellcolor{orange!25}50.00\% (1)\phantom{0} & \cellcolor{orange!25}50.00\% (1)\phantom{0}  & 0.00\% (0) & 0.00\% (0) & 0.00\% (0) \\
R1 &  & 22.69\% (27) & \cellcolor{orange!25}45.38\% (54) & 12.61\% (15) & 5.88\% (7) & 13.45\% (16) \\
R2 &  & 16.44\% (12) & \cellcolor{orange!25}53.42\% (39) & 6.85\% (5) & 6.85\% (5) & 16.44\% (12) \\
R3 &  & 8.00\% (2) & \cellcolor{orange!25}36.00\% (9)\phantom{0}  & 24.00\% (6)\phantom{0}  & 0.00\% (0) & 32.00\% (8)\phantom{0}  \\
R4 &  & 1.35\% (1) & 12.16\% (9)\phantom{0}  & 8.11\% (6) & 8.11\% (6) & \cellcolor{orange!25}70.27\% (52) \\ \hline
\end{tabular}
\begin{tablenotes}
\small
\item[1.] The number of observations is reported in parentheses.
\item[2.] The most frequently observed transitions 
are highlighted.
\end{tablenotes}
\end{threeparttable}
\end{table}

The transition matrix shows that most R1 and R4 players in the ring games remain as the same level in the guessing games. 
Most R2 ring game players, however, only exhibit first-order rationality in the guessing games. 
We do not observe any subjects consistently classified into R3 for both ring and guessing games, possibly because we have relatively low numbers of R3 subjects in either games. 
Overall, there is a relatively high proportion of subjects
(38.23\%) that exhibit the same rationality level
across games.\footnote{GHW report that only 27.3\% of their subjects play at the same level across two families of games.}
Note that in the Robot Treatment, we observe a relatively high proportion (52/293 = 17.74\%) of subjects classified as R4 players in both games,\footnote{In the History Treatment, constant R4 players across games constitute only 6.82\% (20/293) of the subjects. 
See \ref{appendix:extra_table} for the Markov transition
matrix for the History Treatment.} suggesting that subjects in our experiment understand the instruction for robots' decision rules and try to play the best response to such rules.

To test if the high proportion of constant-level players actually results from independent type distributions, we generate 10,000 random samples of 293 pairs of rationality 
levels, independently drawn from the empirical distribution of rationality levels in the Robot Treatment.  
The simulated datasets provide a distribution of the frequency with which a subject plays at the same level in both game families. Furthermore, to establish a baseline for comparison, we utilize the same Monte Carlo simulation and statistical test outlined above to investigate whether the restriction of constant rationality level can be applied to modeling subjects' actions when they face human opponents (choice data) 
in the History Treatment.

Table \ref{tab:h2_simulation} 
reports the simulation results. From the 
table, we can observe that in the Robot Treatment, the 
simulated mean frequency is 32.80\%, with a 95 percent confidence interval ranging from 27.30\% to 38.23\%.
The observed frequency is 38.23\%, rejecting the null hypothesis that the subjects' rationality levels are
independently distributed across games in terms of 
constant rationality levels, at a significance level close to 5\% ($p$-value = 0.058). 
In sharp contrast, the null hypothesis of 
independently distributed rationality levels cannot be rejected 
in the History Treatment, despite the seemingly high proportion 
of constant-level players.
The simulated samples generated from the data in the History Treatment exhibit an average of 40.27\% constant-level
players (95\% CI $=[34.47\%, 45.73\%]$), and the observed frequency in the actual data is 41.30\% 
($p$-value = 0.768).

\begin{table}[htbp!]
\centering
\caption{Constant Level Frequency for the Robot and History Treatment}
\label{tab:h2_simulation}
\renewcommand{\arraystretch}{1.3}
\begin{tabular}{llccc}
\hline
 &  & \multicolumn{3}{c}{Constant Level Frequency} \\ \cline{3-5} 
 &  & Robot Treatment &  & History Treatment \\ \hline
Empirical &  &  &  &  \\
 & Mean: & 38.23\% &  & 41.30\% \\
Simulation &  &  &  &  \\
 & Mean: & 32.80\% &  & 40.27\% \\
 & 95\% CI: & $[27.30\%,\; 38.23\%]$ &  & $[34.47\%,\; 45.73\%]$ \\
 & $p$-value: & \textbf{0.058} &  & \textbf{0.768} \\ \hline
\end{tabular}
\end{table}

Therefore, we conclude that the hypothesis stating that 
individuals exhibit constant rationality levels across games has 
predictive power (though not perfectly accurate) regarding 
experimental subjects' actions under proper belief control.
The sharp contrast between the Robot Treatment and the History Treatment indicates that unifying subjects' beliefs about opponents' depth of reasoning effectively stabilizes the individual revealed rationality level across games.

The constant capacity hypothesis tested 
in this section represents the strictest notion of 
consistency across games, and our findings provide supportive 
evidence for our belief control approach. In \ref{appendix:additional_results}, we explore the consistency of 
rationality levels under different consistency requirements, 
finding that revealed rationality levels remain 
more stable in the 
Robot Treatment than in the History Treatment, and these
results are 
robust across different treatment orders.

\subsection{Cognitive Tests, Secure Actions and 
Strategic Sophistication}
\label{subsection:result_correlation}

\subsubsection{Cognitive Tests and Strategic Sophistication}

If individual strategic sophistication is persistent across games, a natural next question is whether an individual's
performance in other cognitive tests can predict their strategic reasoning ability. 
To explore this, we regress subjects' revealed rationality 
levels on their CRT scores, short-term memory task scores, and farsightedness task scores.

The definitions of the independent variables are as follows:  
\textit{CRT Score} (ranging from 0 to 3) represents the number of correct answers a subject gets in the three CRT questions. 
\textit{Memory Score} (ranging from 0 to 11) is defined as the number of correct answers a subject provides before making the first mistake. 
\textit{Farsightedness} is an indicator variable that equals one 
if a subject chooses to go down at the first move in the farsightedness task (see Section \ref{subsec:cognitive_tests}).
Last, the dependent variable is the individual rationality level (ranging from 0 to 4) revealed in each class of games and each treatment.

\begin{table}[htbp!]
\centering
\begin{threeparttable}
\caption{OLS Regressions for Revealed Rationality Levels}
\label{tab:appendix_regression}
\renewcommand{\arraystretch}{1.3}
\begin{tabular}{rccccc}
\hline
 & \multicolumn{2}{c}{Robot Treatment} &  & \multicolumn{2}{c}{History Treatment} \\ \cline{2-3} \cline{5-6} 
 & Ring Level & Guess Level &  & Ring Level & Guess Level \\ \hline
CRT Score & \phantom{***}0.298*** & \phantom{***}0.566*** &  & \phantom{**}0.239** & \phantom{***}0.461*** \\
 & (0.072) & (0.103) &  & (0.074) & (0.085) \\
Memory Score & 0.026 & 0.030 &  & 0.005 & 0.012 \\
 & (0.036) & (0.032) &  & (0.028) & (0.034) \\
Farsightedness & \phantom{**}0.569** & \phantom{***}0.842*** &  & \phantom{*}0.339* & \phantom{***}0.631*** \\
 & (0.167) & (0.188) &  & (0.167) & (0.165) \\
Constant & \phantom{***}1.058*** & 0.092 &  & \phantom{***}1.078*** & 0.187 \\
 & (0.303) & (0.316) &  & (0.301) & (0.276) \\ \hline
N & 293 & 293 &  & 293 & 293 \\
R-squared & 0.0966 & 0.1788 &  & 0.0556 & 0.1563 \\ \hline
\end{tabular}
\begin{tablenotes}
\small
\item[1.] The standard errors are clustered at the session level.
\item[2.] Significance level: $^*: p<0.05$, $^{**}: p<0.01$, 
$^{***}: p<0.001$.
\end{tablenotes}
\end{threeparttable}
\end{table}

Table \ref{tab:appendix_regression} presents 
the OLS regression results for 
revealed rationality levels. The analysis shows a positive 
correlation between a subject's CRT performance and their 
revealed rationality levels across all game types and 
treatments. Overall, the CRT score is a stronger predictor of 
rationality levels in the guessing games and the Robot Treatment.
In the Robot Treatment, each additional correct answer on the 
CRT is associated with an average increase of 0.298 ($p$-value 
$< 0.001$) in revealed rationality levels for the ring games, 
and 0.566 ($p$-value $< 0.001$) for the guessing games. In 
comparison, in the History Treatment, each additional correct 
answer on the CRT corresponds to a smaller average increase of 
0.239 ($p$-value = 0.002) for the ring games and 0.461 
($p$-value $< 0.001$) for the guessing games---approximately 
80\% of the effect size observed in the Robot Treatment.

In contrast to the previous finding, our results show no significant correlation between short-term memory and strategic sophistication.
The coefficient estimates of \textit{Memory Score} are all below 0.03, and all the corresponding $p$-values are above 0.3.
Notably, these findings are in line with those of GHW, who also observe that CRT scores hold some predictive power over subjects' strategic thinking types, whereas short-term memory capacity does not.

Lastly, an individual's performance on the farsightedness task 
also significantly predicts their revealed rationality level 
across all game types and treatments.
Similar to the CRT score, we observe a stronger correlation between farsightedness and individual rationality levels in the guessing games and in the Robot Treatment.
In the Robot Treatment, a farsighted subject's revealed rationality level is, on average, 0.569 ($p$-value = $ 0.002$) and 
0.842 ($p$-value $< 0.001$) levels higher than that of a myopic subject when playing ring games and guessing games, respectively.
Comparatively, in the History Treatment, a farsighted subject's revealed rationality level is, on average, 0.339 
($p$-value $ = 0.050$) and 0.631 ($p$-value $< 0.001$) levels higher than that of a myopic subject when playing ring games and guessing games, respectively.
Both of these coefficients are smaller in size compared to the estimates reported for the Robot Treatment.
In summary, these results indicate a strong correlation between an important strategic thinking skill in a 
dynamic game---backward induction ability---and the strategic reasoning ability in one-shot interactions.

\subsubsection{Secure Actions in the Ring 
Games\protect\footnote{We thank an anonymous referee for 
suggesting the analysis of secure actions.}}

Another feature of our modified ring games is that, 
except at P4, the secure actions differ from the equilibrium 
actions. This distinction allows us to explore whether players 
opt for secure actions when they have reached their rationality 
capacity. 

\begin{figure}[htbp!]
    \centering
    \includegraphics[width=0.7\linewidth]{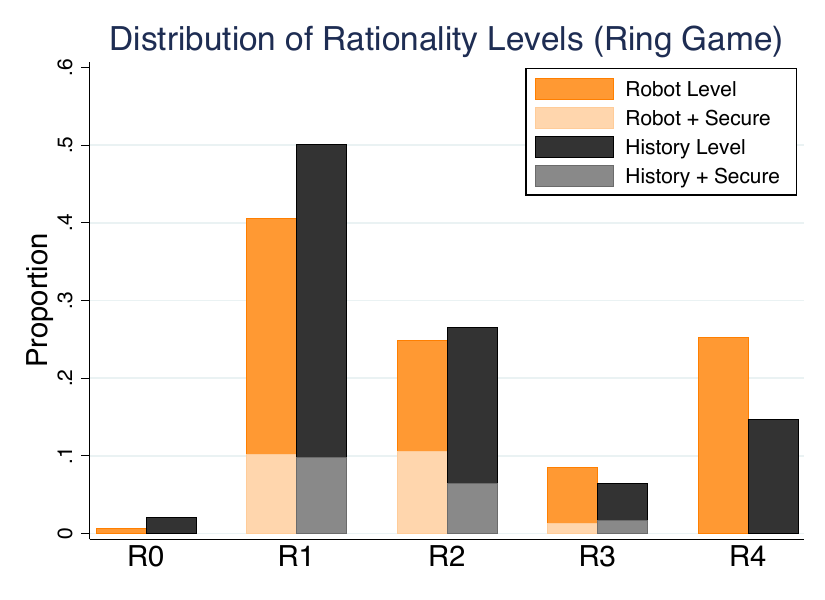}
    \caption{Distribution of Rational Levels with 
    Secure Actions in the Ring Games}
    \label{fig:ring_level_dist_secure}
\end{figure}

In this section, we analyze the behavior of choosing
secure actions by decomposing the 
revealed rationality levels identified from the ring games into 
secure and non-secure types. Specifically, for any rationality 
level $k$, a player is classified as R$k$-Secure (or R$k$-S)
if they exhibit 
rationality level $k$ and choose secure actions in earlier 
positions.\footnote{A player is classified as R3-S if they 
are R3 and choose $(a,a)$ at P1. Similarly, a player is 
classified as R2-S if they are R2 and choose $(a,a)$ at P1 
and $(b,b)$ at P2. A player is classified as R1-S if they 
choose $(a,a)$, $(b,b)$, and $(a,a)$ at P1, P2, and P3, 
respectively.} 
Conversely, a player is classified as R$k$-Non-Secure (or 
R$k$-NS)
if they exhibit rationality level $k$ but do not choose secure 
actions in earlier positions. Based on this classification, players are 
divided into eight possible types: R0, R1-S, R1-NS, 
R2-S, R2-NS, R3-S, R3-NS, and R4. The distributions for the 
Robot and History Treatments are shown in Figure 
\ref{fig:ring_level_dist_secure}.

From this figure, we can first observe that there 
are more secure-type players in the Robot Treatment than in the 
History Treatment. Among the R1 players, 25.2\% are classified 
as R1-S in the Robot Treatment, while 19.7\% are classified as R1-S in the History Treatment. Furthermore, among the R2 players, 42.5\% are R2-S in the Robot Treatment, compared to 24.4\% in the History Treatment. 
This result suggests that instead of betting on risky actions, 
players are more likely to choose a secure action when facing robot players rather than human players.\footnote{Refer to \ref{appendix:additional_results} for the 
joint distribution of rationality levels with secure actions 
across the Robot Treatment and the History Treatment.}

Given this result, we can further explore the behavior of these secure-type players in 
the guessing games. This is an interesting exercise because there is no secure action in 
the guessing games, and one might reasonably hypothesize that secure-type players will exhibit higher rationality levels, as their choice of secure actions in the ring games suggests a degree of deliberate, thoughtful decision-making. However, from the Markov transition matrices in \ref{appendix:extra_table},
we find that, in neither the Robot Treatment nor the History Treatment, for any $k \in 
\{1,2,3\}$, are the transition probabilities between R$k$-S and R$k$-NS significantly 
different.\footnote{To test whether the transition probabilities between 
R$k$-S and R$k$-NS differ, we conduct $\chi^2$ tests, with the null hypothesis that the 
transition probabilities between R$k$-S and R$k$-NS are the same. In the Robot Treatment, 
the $p$-values for R1-S vs. R1-NS, R2-S vs. R2-NS, and R3-S vs. R3-NS are 0.240, 0.338, 
and 0.582, respectively. Similarly, in the History Treatment, the $p$-values for R1-S vs. 
R1-NS, R2-S vs. R2-NS, and R3-S vs. R3-NS are 0.285, 0.211, and 0.476, respectively.} 
This suggests that the existence of secure actions 
is indeed a unique feature of our modified ring games. Given any rationality level, 
choosing secure actions when reaching their rationality capacity does not imply 
significantly different behavior in the guessing games, where secure actions are absent.

\section{Discussions}\label{sec:discussion_new}

\subsection{Validity of Robot Treatment}\label{subsection:robot_validity}

The validity of the Robot Treatment in eliciting individual strategic thinking capacity relies on our Hypothesis 1 (i.e., that individual rationality levels are higher in the Robot Treatment).
An implicit assumption behind this hypothesis is that a subject has an incentive to play at the highest level they can achieve when encountering fully rational opponents playing at their maximum reasoning level.
This statement is trivially true for equilibrium-type subjects, as they know their opponents will play the equilibrium strategy and are able to best respond to it. 
However, for a bounded rational player, this may or may not hold.

If we assume that an iterative reasoning model describes an individual's actual decision-making process, two scenarios explain why a player might only perform $k$ steps of iterative reasoning. 
First, they may incorrectly believe that other players can exhibit (at most) $(k - 1)$th-order of rationality and best respond to that belief. 
Second, they may correctly perceive that other players can exhibit (at least) $k$th-order of rationality but fail to best respond to it. 
While our statement regarding incentive compatibility holds in the first case, it becomes unclear how a bounded rational player would respond when facing opponents with rationality levels above $k$.

Nevertheless, this scenario does not pose a problem under the identification strategy of the revealed rationality approach. 
Notice that a player exhibiting $k$th-order rationality would also exhibit $m$th-order rationality for all $m \leq k$. Thus, a level-$k$ player $i$ who perceives other players as exhibiting at least $k$th-order rationality also perceives them as exhibiting $(k - 1)$th-order rationality. 
That is, the player knows that their robot opponents' strategies will survive $k - 1$ rounds of IEDS. 
Therefore, a payoff-maximizing player $i$ capable of $k$ steps of iterative reasoning will choose a strategy in $R_i^k(\cdot)$, which contains all undominated strategies after $k - 1$ rounds of IEDS. 
Under the revealed rationality approach, player $i$ will then be classified as a $k$th-order revealed-rational player.

Indeed, whether subjects follow the hypothesis and exhibit higher rationality levels when facing fully rational robots is an empirical question. 
In our setting, we have provided supportive evidence for Hypothesis 1 in Section \ref{subsection:result_type}.
However, it remains an open question whether this increased rationality consistently emerges when individuals encounter robot players in other strategic environments.
For instance, in complex games (e.g., Go), individuals might lower their effort and opt for random actions if they perceive highly intelligent robot opponents as unbeatable.
Accordingly, exploring how information about robot opponents may influence people's strategic responses in various settings could deepen our understanding of human-robot interactions, especially as AI increasingly shapes human decision-making processes.

\subsection{Choice of Robot Strategy Instruction}

To elicit individual strategic thinking capacity, our Robot Treatment instructions inform subjects that the computer player is third-order rational (i.e., the computer is rational, knows its opponent is rational, and knows its opponent knows it is rational) to control for their beliefs about the sophisticated robot.
Previous experimental studies have used different approaches to inform subjects about the strategy of a fully rational, equilibrium robot player, such as explaining the concept of equilibrium (e.g., \citealp{costa2006cognition}) or fully disclosing the computer player's exact strategy (e.g., \citealp{meijering2012eye}; \citealp{hanaki2016cognitive}).
However, both approaches may introduce a coaching effect: providing background knowledge about the robot's exact strategy or the concept of equilibrium could directly teach subjects how to play and succeed in the specific game, potentially inflating our estimate of their true strategic thinking capacity.
On the contrary, we describe the robot players' rationality in a multi-layered, recursive manner without providing specific details about their actions \citep{johnson2002detecting}, aiming to reduce the risk of over-coaching while still conveying the robot's strategic sophistication.

Despite our efforts, we acknowledge that our instruction strategy might not fully eliminate the possibility of instruction effects, and some subjects might still be influenced by the way the robot players' rationality is described.
For instance, some subjects might pick up hints on how to apply the logic of IEDS in our dominance-solvable games, thereby enhancing their depth of strategic thinking.
Conversely, others may find the verbal representation of iterative, self-referential logic confusing, which could hinder deeper reasoning.
In future experiments, one could evaluate these effects by running a treatment where subjects read the robot instructions but still play against human opponents, then compare their estimated rationality levels to those in a treatment against humans without robot instructions.

Notably, the goal of our design is to capture individual strategic thinking capacity with respect to iterative reasoning by aligning subjects' beliefs about the robot's higher-order rationality.
As a result, a limitation of our design is that we do not aim to measure how well subjects form beliefs about the overall distribution of the population's strategic reasoning depth and best respond accordingly, which is a key aspect of strategic sophistication in the sense of \cite{stahl1995players}'s ``worldly type.''
An interesting future direction could be to introduce such a ``worldly'' robot player and examine whether subjects could outsmart this strategically sophisticated type when playing against the robot.\footnote{We thank an anonymous referee for encouraging further discussion on our robot strategy instruction.}

\section{Conclusion}\label{sec:conclusion}

This study delves into the cognitive capacity of individuals in strategic interactions. 
To examine their ability to engage in multi-step reasoning, we conduct an experiment designed to elicit and identify each subject's ``rationality bound,'' while controlling for a subject's belief about their opponent's 
depth of reasoning.
Following the revealed rationality approach, we use two classes of dominance solvable games, ring games and guessing games, as the base games in our experiment.  
More importantly, to disentangle the confounding impact of beliefs, we introduce equilibrium-type computer players that are programmed to exhibit infinite order of rationality into the experiment. 
This design allows us to test (1) whether a subject's 
rationality level is (weakly) higher in the Robot Treatment and (2) whether the observed rationality level in the Robot Treatment exhibits any stable pattern across games.

Overall, our results offer compelling evidence that matching subjects with robot players to elicit and identify individual strategic reasoning ability is an effective approach. 
First, subjects exhibit a higher rationality level in the Robot Treatment compared to the History Treatment, supporting the hypothesis that a subject plays at their highest achievable rationality level (i.e., their capacity bound) in the Robot Treatment.
Second, the observed absolute (and relative) order of rationality in the Robot Treatment remains stable across different types of games, a rare finding in previous literature.
Additionally, we find a positive association between a subject's rationality level and their CRT score and backward induction ability, while no significant correlation is observed with short-term memory. 
These findings indicate that strategic reasoning ability may represent an inherent personal characteristic that is distinct from other cognitive abilities and can be reliably inferred from choice data when subjects' beliefs about others are properly controlled.

Considering that the revealed rationality bound identified in the Robot Treatment can serve as a reliable proxy for an individual's strategic thinking ability, we can independently implement dominance-solvable games, such as ring games and guessing games, with human subjects playing against fully rational computer opponents to effectively elicit and identify human players' strategic capacity, either before or after any lab experiment. 
By matching human players with computer players, their revealed strategic sophistication is not confounded by their endogenous beliefs about each other's level of sophistication. 
Furthermore, the robot approach eliminates the need for multiple players to identify a single player's $k$th-order rationality in a game, allowing for an individual task that efficiently elicits and identifies a subject's higher-order rationality. 
Additionally, as the interactions with computer players are independent of the interactions with human players, the two experiences are expected to have minimal influence on each other.
Consequently, the measurement of strategic reasoning ability could remain distinct from the behavioral patterns observed in the main experiment session, thereby avoiding any potential contamination between the two.

Ultimately, we believe that such experiment protocol, particularly the robot approach, has the potential to become a standard tool for measuring a player's actual strategic sophistication, analogous to the usage of the established method (for eliciting risk attitude) in \cite{holt2002risk} but applied to the domain of strategic reasoning. 
By utilizing this tool, we can gain a better understanding of whether non-equilibrium behavior observed in the main experiment can be attributed to bounded strategic thinking capability or other factors, such as non-equilibrium beliefs and social preferences.  

As a final remark, note that our robot strategy instruction is designed by progressively revealing layers of the robot's reasoning. 
By adding or removing these layers, we can introduce a computer player with a higher or lower order of rationality compared to the robot in our experiment, thereby manipulating subjects' beliefs about their opponents' rationality levels.
This flexible, layered structure allows the experimental protocol to be more versatile and applicable to a broader range of contexts, rather than being limited to unifying subjects' beliefs.
Using this instruction strategy, one could experimentally study, for instance, a player's strategic response and its evolution under different distributions of opponents' rationality levels \citep{stahl1993evolution, stahl1995players}.


\newpage
\appendix

\renewcommand{\thesection}{Appendix \Alph{section}}

\section{Additional Tables}
\label{appendix:extra_table}

\setcounter{table}{0}
\renewcommand{\thetable}{A.\arabic{table}}

This appendix includes seven additional tables that 
supplement the analysis in the main text. 
Tables \ref{tab:appendix_ring_raw_count} and 
\ref{tab:appendix_guess_summary} summarize the choices made in 
the ring games and the guessing games, respectively. 
Table \ref{tab:appendix_level_classification} presents the 
distributions of rationality levels under different treatments, 
which are plotted in Figure \ref{fig:level}. Table 
\ref{tab:appendix_ring_level_secure} displays the distributions 
of rationality levels with secure actions, plotted in Figure  
\ref{fig:ring_level_dist_secure}.
The Markov transition matrix for rationality levels in the 
History Treatment is presented in Table \ref{tab:markov_history}.
Finally, Tables \ref{tab:markov_robot_secure} and 
\ref{tab:markov_history_secure} show the Markov transition matrices for rationality levels with secure actions in the Robot Treatment and the History Treatment, respectively.

\bigskip

\begin{table}[htbp!]
\centering
\caption{Number of Observations for Each 
Action Profile in Figure 
\ref{fig:pool_ring_choices} (Ring Games)}
\label{tab:appendix_ring_raw_count}
\renewcommand{\arraystretch}{1.3}
\begin{tabular}{ccccccccccc}
\hline
 &  & \multicolumn{4}{c}{Robot Treatment} &  & \multicolumn{4}{c}{History Treatment} \\ \cline{3-6} \cline{8-11} 
Actions in &  & \multirow{2}{*}{P1} & \multirow{2}{*}{P2} & \multirow{2}{*}{P3} & \multirow{2}{*}{P4} &  & \multirow{2}{*}{P1} & \multirow{2}{*}{P2} & \multirow{2}{*}{P3} & \multirow{2}{*}{P4} \\
G1 and G2 &  &  &  &  &  &  &  &  &  &  \\ \hline
$(a,a)$ &  & 111 & 4 & 40 & 0 &  & 130 & 12 & 49 & 1 \\
$(b,b)$ &  & 2 & 103 & 5 & 0 &  & 1 & 98 & 10 & 0 \\
$(c,c)$ &  & 15 & 4 & 6 & 0 &  & 17 & 5 & 4 & 3 \\
$(a,b)$ &  & 4 & 1 & 61 & 0 &  & 4 & 10 & 73 & 0 \\
$(a,c)$ &  & 53 & 0 & 2 & 2 &  & 72 & 2 & 5 & 1 \\
$(b,a)$ &  & 6 & 57 & 0 & 0 &  & 5 & 79 & 4 & 1 \\
$(b,c)$ &  & 89 & 8 & 0 & 291 &  & 50 & 10 & 0 & 287 \\
$(c,a)$ &  & 11 & 108 & 7 & 0 &  & 14 & 66 & 7 & 0 \\
$(c,b)$ &  & 2 & 8 & 172 & 0 &  & 0 & 11 & 141 & 0 \\ \hline
Total &  & 293 & 293 & 293 & 293 &  & 293 & 293 & 293 & 293 \\ \hline
\end{tabular}
\end{table}

\bigskip

\begin{table}[htbp!]
\centering
\caption{Summary Statistics for Guesses in Guessing Games}
\label{tab:appendix_guess_summary}
\renewcommand{\arraystretch}{1.3}
\begin{tabular}{rccccccc}
\hline
\multicolumn{1}{c}{Treatments} &  & N & Mean & S.D. & Q1 & Median & Q3 \\ \hline
\multicolumn{1}{l}{$p=2/3$} & \multicolumn{1}{l}{} & \multicolumn{1}{l}{} & \multicolumn{1}{l}{} & \multicolumn{1}{l}{} & \multicolumn{1}{l}{} & \multicolumn{1}{l}{} & \multicolumn{1}{l}{} \\
Robot &  & 293 & 28.83 & 24.96 & 1 & 30 & 50 \\
History &  & 293 & 32.28 & 22.40 & 15 & 33 & 45 \\ \hline
\multicolumn{1}{l}{$p=1/3$} & \multicolumn{1}{l}{} & \multicolumn{1}{l}{} & \multicolumn{1}{l}{} & \multicolumn{1}{l}{} & \multicolumn{1}{l}{} & \multicolumn{1}{l}{} & \multicolumn{1}{l}{} \\
Robot &  & 293 & 17.02 & 21.13 & 1 & 12 & 22 \\
History &  & 293 & 19.03 & 19.94 & 5 & 15 & 23 \\ \hline
\multicolumn{1}{l}{$p=1/2$} & \multicolumn{1}{l}{} & \multicolumn{1}{l}{} & \multicolumn{1}{l}{} & \multicolumn{1}{l}{} & \multicolumn{1}{l}{} & \multicolumn{1}{l}{} & \multicolumn{1}{l}{} \\
Robot &  & 293 & 21.50 & 21.41 & 1 & 21 & 30 \\
History &  & 293 & 23.15 & 20.35 & 8 & 24 & 28 \\ \hline
\end{tabular}
\end{table}

\begin{table}[htbp!]
\centering
\caption{Distributions of Rationality Levels in 
Figure \ref{fig:level}}
\label{tab:appendix_level_classification}
\renewcommand{\arraystretch}{1.3}
\begin{tabular}{ccccccccc}
\hline
 &  & \multicolumn{3}{c}{Robot Treatment} &  & \multicolumn{3}{c}{History Treatment} \\ \cline{3-5} \cline{7-9} 
Levels &  & Overall & Ring Game & Guessing Game &  & Overall & Ring Game & Guessing Game \\ \hline
R0 &  & 44 & 2 & 43 &  & 44 & 6 & 39 \\
R1 &  & 149 & 119 & 112 &  & 173 & 147 & 140 \\
R2 &  & 34 & 73 & 32 &  & 47 & 78 & 55 \\
R3 &  & 14 & 25 & 18 &  & 9 & 19 & 12 \\
R4 &  & 52 & 74 & 88 &  & 20 & 43 & 47 \\ \hline
Total &  & 293 & 293 & 293 &  & 293 & 293 & 293 \\ \hline
\end{tabular}
\end{table}

\begin{table}[htbp!]
\centering
\caption{Distributions of Rationality Levels with 
Secure Actions in Figure \ref{fig:ring_level_dist_secure}}
\label{tab:appendix_ring_level_secure}
\renewcommand{\arraystretch}{1.3}
\begin{tabular}{ccccc}
\hline
Levels &  & \begin{tabular}[c]{@{}c@{}}Robot \\ Treatment\end{tabular} &  & \begin{tabular}[c]{@{}c@{}}History \\ Treatment\end{tabular} \\ \hline
R0 &  & 2 &  & 6 \\
\phantom{-S}R1-S &  & 30 &  & 29 \\
\phantom{-NS}R1-NS &  & 89 &  & 118 \\
\phantom{-S}R2-S &  & 31 &  & 19 \\
\phantom{-NS}R2-NS &  & 42 &  & 59 \\
\phantom{-S}R3-S &  & 4 &  & 5 \\
\phantom{-NS}R3-NS &  & 21 &  & 14 \\
R4 &  & 74 &  & 43 \\ \hline
Total &  & 293 &  & 293 \\ \hline
\end{tabular}
\end{table}

\begin{table}[htbp!]
\centering
\begin{threeparttable}
\caption{Markov Transition for Rationality Levels
in the History Treatment}
\label{tab:markov_history}
\renewcommand{\arraystretch}{1.3}
\begin{tabular}{rcccccc}
\hline
\multicolumn{1}{c}{} &  & \multicolumn{5}{c}{Guessing Games} \\
\multicolumn{1}{l}{From $\downarrow$ to $\rightarrow$} &  & R0 & R1 & R2 & R3 & R4 \\ \hline
\multicolumn{1}{l}{Ring Games} &  &  &  &  &  &  \\
R0 &  & 16.67\% (1)\phantom{0} & \cellcolor{orange!25}83.33\% (5)\phantom{0} & 0.00\% (0) & 0.00\% (0) & 0.00\% (0) \\
R1 &  & 15.65\% (23) & \cellcolor{orange!25}58.50\% (86) & 17.69\% (26) & 1.36\% (2) & 6.80\% (10) \\
R2 &  & 15.38\% (12) & \cellcolor{orange!25}43.59\% (34) & 17.95\% (14) & 6.41\% (5) & 16.67\% (13) \\
R3 &  & 10.53\% (2)\phantom{0} & \cellcolor{orange!25}42.11\% (8)\phantom{0} & 26.32\% (5)\phantom{0} & 0.00\% (0) & 21.05\% (4)\phantom{0} \\
R4 &  & 2.33\% (1) & 16.28\% (7)\phantom{0} & 23.26\% (10) & 11.63\% (5)\phantom{0} & \cellcolor{orange!25}46.51\% (20) \\ \hline
\end{tabular}
\begin{tablenotes}
\small
\item[1.] The number of observations is reported in parentheses.
\item[2.] The most frequently observed transitions 
are highlighted.
\end{tablenotes}
\end{threeparttable}
\end{table}

\begin{table}[htbp!]
\centering
\begin{threeparttable}
\caption{Markov Transition for Rationality Levels in the Robot Treatment}
\label{tab:markov_robot_secure}
\renewcommand{\arraystretch}{1.3}
\begin{tabular}{rcccccc}
\hline
\multicolumn{1}{c}{} &  & \multicolumn{5}{c}{Guessing Games} \\
\multicolumn{1}{l}{From $\downarrow$ to $\rightarrow$} &  & R0 & R1 & R2 & R3 & R4 \\ \hline
\multicolumn{1}{l}{Ring Games} &  &  &  &  &  &  \\
R0 &  & \cellcolor{orange!25}50.00\% (1)\phantom{0} & \cellcolor{orange!25}50.00\% (1)\phantom{0} & 0.00\% (0) & 0.00\% (0) & 0.00\% (0) \\
R1-S &  & 20.00\% (6)\phantom{0} & \cellcolor{orange!25}60.00\% (18) & 6.67\% (2) & 0.00\% (0) & 13.33\% (4)\phantom{0} \\
R1-NS &  & 23.60\% (21) & \cellcolor{orange!25}40.45\% (36) & 14.61\% (13) & 7.87\% (7) & 13.48\% (12) \\
R2-S &  & 16.13\% (5)\phantom{0} & \cellcolor{orange!25}61.29\% (19) & 6.45\% (2) & 9.68\% (3) & 6.45\% (2) \\
R2-NS &  & 16.67\% (7)\phantom{0} & \cellcolor{orange!25}47.62\% (20) & 7.14\% (3) & 4.76\% (2) & 23.81\% (10) \\
R3-S &  & 0.00\% (0) & 25.00\% (1)\phantom{0} &\cellcolor{orange!25} 50.00\% (2)\phantom{0} & 0.00\% (0) & 25.00\% (1)\phantom{0} \\
R3-NS &  & 9.52\% (2) & \cellcolor{orange!25}38.10\% (8)\phantom{0} & 19.05\% (4)\phantom{0} & 0.00\% (0) & 33.33\% (7)\phantom{0} \\
R4 &  & 1.35\% (1) & 12.16\% (9)\phantom{0} & 8.11\% (6) & 8.11\% (6) & \cellcolor{orange!25}70.27\% (52) \\ \hline
\end{tabular}
\begin{tablenotes}
\small
\item[1.] The number of observations is reported in parentheses.
\item[2.] The most frequently observed transitions 
are highlighted.
\end{tablenotes}
\end{threeparttable}
\end{table}

\begin{table}[htbp!]
\centering
\begin{threeparttable}
    \caption{Markov Transition for Rationality Levels in the    History Treatment}
\label{tab:markov_history_secure}
\renewcommand{\arraystretch}{1.3}
\begin{tabular}{rcccccc}
\hline
\multicolumn{1}{c}{} &  & \multicolumn{5}{c}{Guessing Games} \\
\multicolumn{1}{l}{From $\downarrow$ to $\rightarrow$} &  & R0 & R1 & R2 & R3 & R4 \\ \hline
\multicolumn{1}{l}{Ring Games} &  &  &  &  &  &  \\
R0 &  & 16.67\% (1)\phantom{0} & \cellcolor{orange!25}83.33\% (5)\phantom{0} & 0.00\% (0) & 0.00\% (0) & 0.00\% (0) \\
R1-S &  & 20.69\% (6)\phantom{0} & \cellcolor{orange!25}65.52\% (19) & 6.90\% (2) & 3.45\% (1) & 3.45\% (1) \\
R1-NS &  & 14.41\% (17) & \cellcolor{orange!25}56.78\% (67) & 20.34\% (24) & 0.85\% (1) & 7.63\% (9) \\
R2-S &  & 21.05\% (4)\phantom{0} & \cellcolor{orange!25}52.63\% (10) & 0.00\% (0) & 5.26\% (1) & 21.05\% (4)\phantom{0} \\
R2-NS &  & 13.56\% (8)\phantom{0} & \cellcolor{orange!25}40.68\% (24) & 23.73\% (14) & 6.78\% (4) & 15.25\% (9)\phantom{0} \\
R3-S &  & 20.00\% (1)\phantom{0} & \cellcolor{orange!25}40.00\% (2)\phantom{0} & \cellcolor{orange!25}40.00\% (2)\phantom{0} & 0.00\% (0) & 0.00\% (0) \\
R3-NS &  & 7.14\% (1) & \cellcolor{orange!25}42.86\% (6)\phantom{0} & 21.43\% (3)\phantom{0} & 0.00\% (0) & 28.57\% (4)\phantom{0} \\
R4 &  & 2.33\% (1) & 16.28\% (7)\phantom{0} & 23.26\% (10) & 11.63\% (5)\phantom{0} & \cellcolor{orange!25}46.51\% (20) \\ \hline
\end{tabular}
\begin{tablenotes}
\small
\item[1.] The number of observations is reported in parentheses.
\item[2.] The most frequently observed transitions 
are highlighted.
\end{tablenotes}
\end{threeparttable}
\end{table}

\newpage
\section{Additional Analysis}
\label{appendix:additional_results}

In this appendix, we provide 
additional analysis to complement the main text.
In the first section, we report the joint distributions of 
rationality levels across treatments for the ring games and the 
guessing games, which supplement the analysis in Section 
\ref{subsection:result_type} of the main text. Next, 
we analyze the empirical best responses in the ring games in the History 
Treatment. In the following sections, we examine the consistency of 
rationality levels across games under weaker notions of 
consistency compared to the one analyzed in Section \ref{subsection:result_constant} of the main text.

\setcounter{table}{0}
\renewcommand{\thetable}{B.\arabic{table}}
\setcounter{figure}{0}
\renewcommand{\thefigure}{B.\arabic{figure}}

\subsection*{Joint Distribution of Rationality Levels}

\begin{table}[htbp!]
\centering
\begin{threeparttable}
\caption{Joint Distribution of Rationality Levels in 
the Ring Games}
\label{tab:appendix_joint_ring}
\renewcommand{\arraystretch}{1.3}
\begin{tabular}{rcccccc}
\hline
\multicolumn{1}{c}{} & \multicolumn{1}{l}{} & \multicolumn{5}{c}{History Treatment} \\
\multicolumn{1}{l}{\textbf{Ring Games}} & \multicolumn{1}{l}{} & R0 & R1 & R2 & R3 & R4 \\ \hline
\multicolumn{1}{l}{Robot Treatment} &  &  &  &  &  &  \\
R0 &  & 0.34\% (1) & 0.00\% (0) & 0.34\% (1)\phantom{0} & 0.00\% (0) & 0.00\% (0) \\
R1 &  & 1.37\% (4) & 27.99\% (82) & 8.53\% (25) & 1.37\% (4) & 1.37\% (4) \\
R2 &  & 0.34\% (1) & 12.29\% (36) & 8.87\% (26) & 2.05\% (6) & 1.37\% (4) \\
R3 &  & 0.00\% (0) & \phantom{0}3.41\% (10) & 2.39\% (7)\phantom{0} & 0.68\% (2) & 2.05\% (6) \\
R4 &  & 0.00\% (0) & \phantom{0}6.48\% (19) & 6.48\% (19) & 2.39\% (7) & \phantom{0}9.90\% (29) \\ \hline
\end{tabular}
\begin{tablenotes}
\small
\item[1.] The number of observations is reported in parentheses.
\end{tablenotes}
\end{threeparttable}
\end{table}

\begin{table}[htbp!]
\centering
\begin{threeparttable}
\caption{Joint Distribution of Rationality Levels in 
the Guessing Games}
\label{tab:appendix_joint_guessing}
\renewcommand{\arraystretch}{1.3}
\begin{tabular}{rcccccc}
\hline
\multicolumn{1}{c}{} & \multicolumn{1}{l}{} & \multicolumn{5}{c}{History Treatment} \\
\multicolumn{1}{l}{\textbf{Guessing Games}} & \multicolumn{1}{l}{} & R0 & R1 & R2 & R3 & R4 \\ \hline
\multicolumn{1}{l}{Robot Treatment} &  &  &  &  &  &  \\
R0 &  & 8.53\% (25) & \phantom{0}6.14\% (18) & 0.00\% (0)\phantom{0} & 0.00\% (0) & 0.00\% (0) \\
R1 &  & 4.10\% (12) & 27.99\% (82) & 6.14\% (18) & 0.00\% (0) & 0.00\% (0) \\
R2 &  & 0.00\% (0)\phantom{0} & \phantom{0}4.44\% (13) & 4.44\% (13) & 1.37\% (4) & 0.68\% (2) \\
R3 &  & 0.34\% (1)\phantom{0} & 3.07\% (9) & 1.37\% (4)\phantom{0} & 1.02\% (3) & 0.34\% (1) \\
R4 &  & 0.34\% (1)\phantom{0} & \phantom{0}6.14\% (18) & 6.83\% (20) & 1.71\% (5) & 15.02\% (44) \\ \hline
\end{tabular}
\begin{tablenotes}
\small
\item[1.] The number of observations is reported in parentheses.
\end{tablenotes}
\end{threeparttable}
\end{table}

In this section, we present the joint 
distributions of rationality levels across both the Robot 
Treatment and the History Treatment for both games. The joint
distributions for the ring games and the guessing games 
are shown in Table \ref{tab:appendix_joint_ring} and 
Table \ref{tab:appendix_joint_guessing}, respectively.
Overall, 72 percent of subjects (211/293) 
exhibit (weakly) higher rationality levels in the Robot Treatment compared to the History Treatment 
in both games. In contrast, fewer than four percent of subjects (11/293) consistently exhibit strictly lower rationality levels in the Robot Treatment across games.
We further conduct a Wilcoxon signed-rank test to examine whether the subjects' rationality levels in the Robot Treatment are significantly greater than those in the History Treatment. Consistent with Hypothesis 1, we observe higher rationality levels in the Robot Treatment (Wilcoxon test $p$-value < 0.0001 for both the ring games and guessing games).

Additionally, Table \ref{tab:appendix_joint_ring_secure} reports the 
joint distribution of rationality levels for the ring 
games using the classification rule introduced in 
Section \ref{subsection:result_correlation}. From the table, 
we observe that secure-type players identified in the History 
Treatment are more likely to be classified as secure-type in the Robot Treatment as well.
For example, 20 of the 29 R1-S types in the History Treatment 
are classified as either R1-S or R2-S types in the Robot Treatment. 
In contrast, among the 118 R1-NS players in the History Treatment, 
only 22 are classified as secure types in the Robot Treatment. 
This finding provides suggestive evidence for the stability of secure types.

\begin{table}[htbp!]
\centering
\caption{Joint Distribution of Rationality Levels with 
Secure Action in the Ring Games}
\label{tab:appendix_joint_ring_secure}
\renewcommand{\arraystretch}{1.3}
\begin{adjustbox}{width=\columnwidth,center}
\begin{threeparttable}
\begin{tabular}{rccccccccc}
\hline
\multicolumn{1}{c}{} &  & \multicolumn{8}{c}{History Treatment} \\
\multicolumn{1}{l}{\textbf{Ring Games}} &  & R0 & R1-S & R1-NS & R2-S & R2-NS & R3-S & R3-NS & R4 \\ \hline
\multicolumn{1}{l}{Robot Treatment} &  &  &  &  &  &  &  &  &  \\
R0 &  & 0.34\% (1) & 0.00\% (0) & 0.00\% (0) & 0.00\% (0) & 0.34\% (1) & 0.00\% (0) & 0.00\% (0) & 0.00\% (0) \\
R1-S &  & 0.00\% (0) & \phantom{0}4.44\% (13) & \phantom{0}3.75\% (11) & 1.02\% (3) & 0.68\% (2) & 0.00\% (0) & 0.34\% (1) & 0.00\% (0) \\
R1-NS &  & 1.37\% (4) & 2.05\% (6) & 17.75\% (52) & 2.39\% (7) & \phantom{0}4.44\% (13) & 0.00\% (0) & 1.02\% (3) & 1.37\% (4) \\
R2-S &  & 0.00\% (0) & 2.39\% (7) & 2.73\% (8) & 1.71\% (5) & 2.39\% (7) & 1.02\% (3) & 0.00\% (0) & 0.34\% (1) \\
R2-NS &  & 0.34\% (1) & 0.34\% (1) & \phantom{0}6.83\% (20) & 0.34\% (1) & \phantom{0}4.44\% (13) & 0.34\% (1) & 0.68\% (2) & 1.02\% (3) \\
R3-S &  & 0.00\% (0) & 0.00\% (0) & 1.02\% (3) & 0.00\% (0) & 0.34\% (1) & 0.00\% (0) & 0.00\% (0) & 0.00\% (0) \\
R3-NS &  & 0.00\% (0) & 0.00\% (0) & 2.39\% (7) & 0.34\% (1) & 1.71\% (5) & 0.00\% (0) & 0.68\% (2) & 2.05\% (6) \\
R4 &  & 0.00\% (0) & 0.68\% (2) & \phantom{0}5.80\% (17) & 0.68\% (2) & \phantom{0}5.80\% (17) & 0.34\% (1) & 2.05\% (6) & \phantom{0}9.90\% (29) \\ \hline
\end{tabular}
\begin{tablenotes}
\small
\item[1.] The number of observations is reported in parentheses.
\end{tablenotes}
\end{threeparttable}
\end{adjustbox}

\end{table}

\subsection*{Empirical Best Responses in the Ring Games in the History Treatment}

Figure \ref{fig:pool_ring_choices} reveals a subtle but interesting 
pattern: a higher proportion of subjects choose equilibrium actions in G2 but secure 
actions in G1 in the History Treatment compared to the Robot Treatment. 
We analyze the empirical best responses in the ring games from the History Treatment to explore whether these responses may provide a rationale for this behavior.

\begin{figure}[htbp!]
    \centering
    \includegraphics[width=\linewidth]{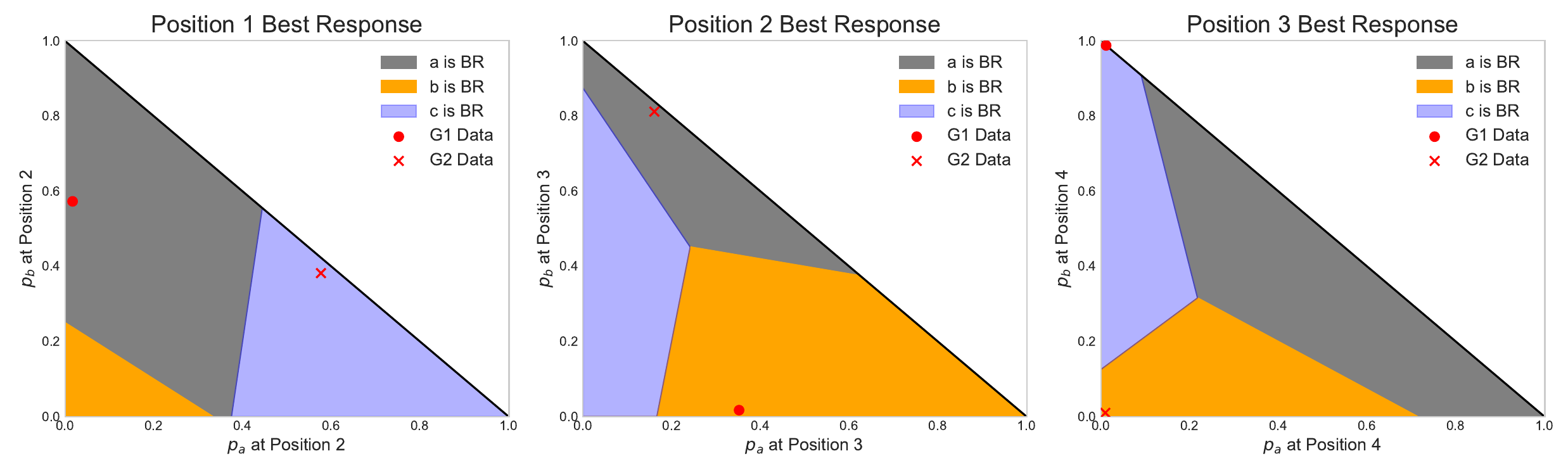}
    \caption{Best response regions in the ring games at Position 1 (left), Position 2
    (middle), and Position 3 (right). The empirical choice frequencies of the player at 
    the next position in the Robot Treatment are marked with circles for G1 and crosses 
    for G2.}
    \label{fig:ring_empirical_br}
\end{figure}

Figure \ref{fig:ring_empirical_br} shows the best response function for each position 
in the ring games, with Position 4 omitted due to its strictly dominant strategy ($b$ 
in G1 and $c$ in G2). Since the payoff at position $n$ depends only on the choice made 
by the player at position $n$ and the choice at position $n+1$, each panel of Figure 
\ref{fig:ring_empirical_br} illustrates the best response regions within the 
probability simplex of the next player's choice probabilities. 
Take Position 1 as an example. The lower left (orange) area corresponds to the scenario 
where the player at Position 2 chooses $a$ and $b$ with low probabilities, making $b$ the 
best response for the player at Position 1.

Combined with the empirical choice frequencies from the Robot Treatment 
(marked with circles for G1 and crosses for G2), we observe that in the History 
Treatment, the empirical best response at Position 1 is $(a,c)$, while at Position 2, 
it is $(b,a)$. This suggests that at Positions 1 and 2, choosing secure actions in G1 
but equilibrium actions in G2 indeed corresponds to the 
empirical best response. However, this behavior is not the empirical best response at 
Position 3. Given that nearly all subjects choose equilibrium actions at Position 4, 
the empirical best response there is $(c,b)$, the equilibrium actions.

Given this result, we can further decompose the revealed rationality levels identified 
from the ring games in the History Treatment by introducing two additional types: R2-BR 
and R3-BR. These players are classified as R2 and R3, respectively, and they choose 
secure actions in earlier positions in G1 and equilibrium actions in earlier positions in 
G2. Among the 59 R2-NS subjects in the History Treatment, 15 are now classified as R2-BR. 
Additionally, 10 out of the 14 R3-NS subjects in the History Treatment are classified as 
R3-BR.

These empirical best responders are indeed highly rational, as 11 of the 15 R2-BR players 
are classified as R3 or R4 in the ring games in the Robot Treatment. Similarly, 7 out of 
10 R3-BR players are either R3 or R4 in the Robot Treatment. Although the number of 
observations is limited, it suggests that some highly rational 
players would deliberately deviate from equilibrium actions in the History Treatment to best 
respond to the empirical data.

\subsection*{Constant Ranking of Rationality Levels}

In this section, we evaluate the consistency of 
rationality levels across games under a weaker notion of 
consistency compared to the constant capacity hypothesis 
(Hypothesis 2). Rather than assessing whether rationality levels 
remain constant across games, we test whether the ranking of 
players, in terms of rationality levels, stays the same. In 
other words, we examine whether playing against robots provides 
a \emph{relative} measure of rationality levels. Formally, we 
test the following hypothesis.

\bigskip
\noindent\textbf{Hypothesis 3 (Constant Ranking of 
Capacity).} For every $i, j \in N$, $k_i(\gamma,\; \mbox{Robot})
\geq k_j(\gamma,\; \mbox{Robot}) $ for some $\gamma$ 
implies $k_i(\gamma',\; \mbox{Robot}) \geq k_j(\gamma',\; \mbox{Robot}) $ for all $\gamma' \in \Gamma$.
\bigskip

To test this hypothesis, we follow GHW in defining the 
\textit{switch frequency}, \textit{non-switch frequency} and 
the \textit{switch ratio}.
The \textit{switch frequency} represents the proportion of player pairs in which the player who exhibits a strictly higher level in one game becomes the player with a strictly lower level in another game. 
On the other hand, the \textit{non-switch frequency} corresponds to the proportion of player pairs in which the player with a strictly higher level in one game maintains that higher level in another game.\footnote{The sum of the switch frequency and non-switch frequency may not be one since the paired players who exhibit the same level in one game are excluded.} 
The \textit{switch ratio} is calculated by dividing the switch frequency by the non-switch frequency.  
If the relative rationality levels are preserved across games, the switch ratio will be zero. 
Alternatively, if the rationality levels are independently drawn, we expect to observe a switch ratio of one.

Table \ref{tab:switch_robot} reports the switch frequency, non-switch frequency, and switch ratio observed in the actual data and computed under the null hypothesis of independently distributed rationality levels. 
The results presented provide strong evidence of
stable rankings of individual rationality levels.
Our pooled data show that non-switching occurs three times more frequently than switching in the Robot Treatment (Non-switching: 41.30\%; Switching: 12.28\%). 
The switch ratio of 0.30, derived from the switch and non-switch frequencies, is lower than any switch ratio obtained from our 10,000-sample simulated data.  
Additionally, these results consistently hold across different treatment orders.
Whether the Robot Treatment is played first or second, the observed switch ratios remain around 0.30, both of which are lower than any switch ratio obtained from the simulated data. 
Consequently, we reject the null hypothesis of independently distributed levels in terms of relative rationality depths, with a $p$-value less than 0.0001.

\begin{table}[htbp!]
\centering
\caption{Switch Ratio for the Robot and History Treatment}
\label{tab:switch_robot}
\begin{adjustbox}{width=\columnwidth,center}
\begin{tblr}{
  column{2} = {c},
  column{3} = {c},
  column{5} = {c},
  column{6} = {c},
  column{8} = {c},
  column{9} = {c},
  cell{1}{1} = {c},
  cell{1}{2} = {c=2}{c},
  cell{1}{5} = {c=2}{c},
  cell{1}{7} = {c},
  cell{1}{8} = {c=2}{c},
  cell{2}{1} = {c},
  hline{1,3,14} = {-}{},
  hline{2} = {2-3,5-6,8-9}{},
}
                                 & Pooled Data ($n = 293$)           &                     &  & RH Order ($n = 136$)          &                     &  & HR Order ($n = 157$)          &                     \\
{Ring Game vs. \\ Guessing Game} & {Empirical\\ Data} & {Null\\ Hypothesis} &  & {Empirical\\ Data} & {Null\\ Hypothesis} &  & {Empirical\\ Data} & {Null\\ Hypothesis} \\
Robot Treatment                  &                    &                     &  &                    &                     &  &                    &                     \\
\;\; Switch frequency:                & 12.28\%             & 22.56\%              &  & 11.94\%             & 19.77\%              &  & 12.45\%             & 24.05\%              \\
\;\; Non-switch frequency:            & 41.30\%             & 22.58\%              &  & 37.69\%             & 19.84\%              &  & 42.25\%             & 24.06\%              \\
\;\; Switch ratio:                    & 0.30               & 1.01                &  & 0.32               & 1.03                &  & 0.29               & 1.02                \\
\;\; $p$-value:                         & $<0.0001$         &                     &  & $<0.0001$          &                     &  & $<0.0001$          &                     \\
                                 &                    &                     &  &                    &                     &  &                    &                     \\
History Treatment                &                    &                     &  &                    &                     &  &                    &                     \\
\;\; Switch frequency:                & 12.89\%             & 17.84\%              &  & 11.05\%             & 21.30\%              &  & 14.77\%             & 14.53\%              \\
\;\; Non-switch frequency:            & 34.47\%             & 17.87\%              &  & 40.26\%             & 21.28\%              &  & 28.12\%             & 14.52\%              \\
\;\; Switch ratio:                    & 0.37               & 1.01                &  & 0.27               & 1.03                &  & 0.53               & 1.04                \\
\;\; $p$-value:                         & $<0.0001$         &                     &  & $<0.0001$          &                     &  & 0.020              &                     
\end{tblr}
\end{adjustbox}
\end{table}

We also calculate the switch and non-switch frequencies in the History Treatment to investigate whether the rankings of individual rationality levels remain stable when subjects' beliefs about others' rationality depths are not controlled. 
In the History Treatment, the null hypothesis of independently distributed levels in terms of relative rationality depths is also rejected ($p$-value $ < 0.0001$), with the pooled data showing switch and non-switch frequencies of 12.89\% and 34.47\%, respectively, resulting in a switch ratio of 0.37. 
However, it is noteworthy that the switch ratio in the History Treatment is 23\% higher than that in the Robot Treatment, and this difference increases to 66\% when focusing solely on the Robot and History Treatments that are played first by subjects (Robot: 0.32; History: 0.53).\footnote{We conduct a statistical comparison by contrasting a switch ratio of 0.32 with the switch ratios obtained from 10,000 random samples of independently drawn levels from the empirical distribution of rationality levels in the History Treatment under HR Order. Our analysis reveals that, when focusing exclusively on the data from treatments played first, we can reject the null hypothesis that the observed rationality levels in the Robot Treatment are drawn from the same distribution of levels as in the History Treatment, in terms of switch ratios 
($p$-value = $0.026$).}

This result primarily stems from the significantly higher 
non-switch frequency in the Robot Treatment compared to the 
History Treatment. These findings suggest that unifying 
subjects' beliefs about their opponents' strategic reasoning 
capabilities significantly improves the stability of individual 
rationality levels across games. This indicates that strategic 
reasoning ability may be an inherent personal characteristic, 
which can be inferred from choice data when participants 
interact with robot players.

\subsection*{Constant Ranking of Games}

In this section, we use the same analysis method 
to evaluate whether playing against robot players can provide a 
measure of game difficulty based on players' depth of reasoning. 
Specifically, we assess whether the ranking of games, in terms 
of a player's rationality level, remains consistent across 
different players. This is formalized in the following 
hypothesis.

\bigskip
\noindent\textbf{Hypothesis 4 (Constant Ranking of Games).} For 
every $\gamma, \gamma' \in \Gamma$, $k_i(\gamma, \;
\mbox{Robot}) \geq k_i(\gamma', \;
\mbox{Robot})$ for some player $i$ implies $k_j(\gamma, \;
\mbox{Robot}) \geq k_j(\gamma', \;
\mbox{Robot})$ for all $j\in N$.
\bigskip

Similar to the previous analysis, we define the 
\emph{change-in-same-direction frequency}, 
\emph{change-in-opposite-directions frequency} and 
\emph{opposite/same ratio}. 
The \textit{change-in-same-direction frequency} represents the proportion of player pairs in which both players exhibit a strictly higher level in the same game.
In contrast, the \textit{change-in-opposite-directions frequency} refers to the proportion of player pairs in which the two players exhibit a strictly higher rationality level in different games.\footnote{The sum of the change-in-same-direction frequency and change-in-opposite-directions frequency may not be one, as a pair of players is excluded if one of them exhibits the same level across games.} The \textit{opposite/same ratio} is calculated by dividing the 
change-in-opposite-directions frequency by the change-in-same-direction frequency.
If the ranking of games remains constant across players, the opposite/same ratio would be zero. Conversely, if rationality levels are independently distributed, we would expect the opposite/same ratio to be one

\begin{table}[htbp!]
\centering
\caption{Opposite/same Ratio for the Robot and History Treatment}
\label{tab:difficulty_robot}
\begin{adjustbox}{width=\columnwidth,center}
\begin{tblr}{
  column{2} = {c},
  column{3} = {c},
  column{5} = {c},
  column{6} = {c},
  column{8} = {c},
  column{9} = {c},
  cell{1}{1} = {c},
  cell{1}{2} = {c=2}{c},
  cell{1}{5} = {c=2}{c},
  cell{1}{7} = {c},
  cell{1}{8} = {c=2}{c},
  cell{2}{1} = {c},
  hline{1,3,14} = {-}{},
  hline{2} = {2-3,5-6,8-9}{},
}
                                   & Pooled Data ($n = 293$)          &                     &  & RH Order ($n = 136$)          &                     &  & HR Order ($n = 157$)          &                     \\
{Ring Game vs. \\ Guessing Game}   & {Empirical\\ Data} & {Null\\ Hypothesis} &  & {Empirical\\ Data} & {Null\\ Hypothesis} &  & {Empirical\\ Data} & {Null\\ Hypothesis} \\
Robot Treatment                    &                    &                     &  &                    &                     &  &                    &                     \\
\;\; Change in opposite direction: & 17.50\%             & 22.58\%              &  & 18.69\%             & 19.80\%              &  & 16.45\%             & 24.05\%              \\
\;\; Change in same direction:     & 20.58\%             & 22.58\%              &  & 20.20\%             & 19.81\%              &  & 20.78\%             & 24.06\%              \\
\;\; Opposite/same ratio:               & 0.85               & 1.00                &  & 0.93               & 1.00                &  & 0.79               & 1.00                \\
\;\; $p$-value:                           & $<0.0001$  &                     &  & 0.078              &                     &  & 0.0004             &                     \\
                                   &                    &                     &  &                    &                     &  &                    &                     \\
History Treatment                  &                    &                     &  &                    &                     &  &                    &                     \\
\;\; Change in opposite direction: & 16.26\%             & 17.84\%              &  & 17.08\%             & 21.27\%              &  & 15.63\%             & 14.51\%              \\
\;\; Change in same direction:     & 18.12\%             & 17.84\%              &  & 18.21\%             & 21.28\%              &  & 17.81\%             & 14.50\%              \\
\;\; Opposite/same ratio:               & 0.90               & 1.00                &  & 0.94               & 1.00                &  & 0.88               & 1.00                \\
\;\; $p$-value:                           & 0.002              &                     &  & 0.110              &                     &  & 0.022              &                     
\end{tblr}
\end{adjustbox}
\end{table}

Table \ref{tab:difficulty_robot} reports the 
change-in-same-direction frequency, change-in-opposite-directions frequency, and the opposite/same ratio computed based on actual data and simulated data generated from independently-drawn levels. 
In the Robot Treatment, the frequency with which two paired players change their rationality levels in the same direction (20.58\%) is 3 percentage point higher than the frequency of changing in the opposite directions (17.50\%), as shown in Table \ref{tab:difficulty_robot} (the column of Pooled Data). 
The observed opposite/same ratio of 0.85 significantly deviates from the mean of the simulated datasets (1.00 with a 95 percent confidence interval of 0.96 to 1.01), leading to the rejection of the null hypothesis of independently distributed levels in terms of the ranking of games ($p$-value $< 0.0001$).
This result remains robust regardless of the order of treatments, although it only reaches marginal significance when the analysis is limited to the subjects who played the Robot Treatment first 
(RH Order: $p$-value = $ 0.078$; HR Order: $p$-value = $0.0004$).
Consequently, our findings suggest that an individual's strategic reasoning level, when beliefs are properly controlled, can serve as a reliable proxy for the relative complexity or difficulty of a game.

Finally, in the History Treatment, we find a similar result but with weaker evidence.
The simulated datasets generated from the History Treatment data yield a mean opposite/same ratio of 1.00, with a 95 percent confidence interval of 0.95 to 1.01. 
The actual ratio of 0.90, which is 6\% higher than that in the Robot Treatment, still rejects the null hypothesis of independently distributed levels with a significance level of 
$p$-value = $ 0.002$. 
However, this result becomes less robust when considering the order of treatments. 
We cannot reject the null hypothesis when examining only the subjects who played against robots before playing against human choice data 
(RH Order: $p$-value = $  0.110$; HR Order: $p$-value = $ 0.022$). 
Without controlling for a subject's belief about their opponents' strategic thinking abilities, the observed rationality level could reflect either the complexity of the environment or how a subject believes others would perceive the complexity of the environment, and thus has weaker predictive power on other players' (revealed) rationality level.

\end{document}